\documentclass{emulateapj}

\usepackage{appendix}
\usepackage{apjfonts}
\usepackage{lscape}

\newcommand {\etal} {et~al.~}
\def \spose#1{\hbox  to 0pt{#1\hss}}  
\newcommand {\lta} {\mathrel{\spose{\lower 3pt\hbox{$\sim$}}\raise  2.0pt\hbox{$<$}}}
\newcommand {\gta} {\mathrel{\spose{\lower  3pt\hbox{$\sim$}}\raise 2.0pt\hbox{$>$}}}
\def \ion#1#2{#1{\footnotesize{#2}}\relax} 
\newcommand {\ha}  {\ifmmode H\alpha \else H$\alpha $ \fi} 
\newcommand {\hi} {\ion{H}{I}} 

\newcommand {\kmsmpc} {\>{\rm km}\,{\rm s}^{-1}\,{\rm Mpc}^{-1}}
\newcommand {\kms} {\ifmmode  \,\rm km\,s^{-1} \else $\,\rm km\,s^{-1}  $ \fi }
\newcommand {\kpc} {\ifmmode  {\rm kpc}  \else ${\rm  kpc}$ \fi  }  
\newcommand {\Msun} {\ifmmode M_{\odot} \else $M_{\odot}$ \fi} 
\newcommand {\Lsun} {\ifmmode L_{\odot} \else $L_{\odot}$ \fi} 
\newcommand {\magarc}{\ifmmode {{{{\rm mag}~{\rm arcsec}}^{-2}}}
             \else {{{mag}$~${arcsec}$^{-2}$}}
             \fi}

\newcommand {\LCDM} {\ifmmode \Lambda{\rm CDM} \else $\Lambda{\rm CDM}$ \fi}
\newcommand {\OmegaM} {\ifmmode \Omega_{\rm m} \else $\Omega_{\rm m}$ \fi} 
\newcommand {\OmegaL} {\ifmmode \Omega_{\rm \Lambda} \else $\Omega_{\rm \Lambda}$\fi} 
\newcommand {\Deltavir} {\ifmmode \Delta_{\rm vir} \else $\Delta_{\rm vir}$ \fi}

\newcommand {\aac} {\ifmmode \alpha_{\rm AC} \else $\alpha_{\rm AC}$ \fi} 
\newcommand {\rs} {\ifmmode r_{\rm s} \else $r_{\rm s}$ \fi} 
\newcommand {\Rvir} {\ifmmode R_{\rm vir} \else $R_{\rm vir}$ \fi}
\newcommand {\Vvir} {\ifmmode V_{\rm  vir} \else  $V_{\rm vir}$  \fi} 
\newcommand {\Mvir} {\ifmmode M_{\rm  vir} \else $M_{\rm  vir}$ \fi}  
\newcommand {\Jvir} {\ifmmode J_{\rm vir} \else $J_{\rm vir}$ \fi} 

\newcommand {\lamgal} {\ifmmode \lambda_{\rm gal}  \else $\lambda_{\rm gal}$ \fi} 
\newcommand {\lam} {\ifmmode \lambda  \else $\lambda$ \fi} 
\newcommand {\lambar} {\ifmmode \bar{\lambda}  \else  $\bar{\lambda}$  \fi}  

\newcommand {\dd} {\ifmmode {\rm d} \else ${\rm d}$ \fi} 

\newcommand {\alpham} {\ifmmode \alpha_{\rm m} \else $\alpha_{\rm m}$ \fi} 
\newcommand {\fbar} {\ifmmode f_{\rm bar} \else $f_{\rm bar}$ \fi} 
\newcommand {\RI} {\ifmmode R_{I} \else $R_{I}$ \fi} 
\newcommand {\Rs} {\ifmmode R_{*} \else $R_{*}$ \fi}  
\newcommand {\Rd} {\ifmmode R_{\rm d} \else $R_{\rm d}$ \fi}  
\newcommand {\Reff} {\ifmmode R_{\rm eff} \else $R_{\rm  eff}$ \fi} 
\newcommand {\rb} {\ifmmode r_{\rm b}  \else $r_{\rm b}$ \fi}
\newcommand {\ri} {\ifmmode r_{\rm i}  \else $r_{\rm i}$ \fi}
\newcommand {\rf} {\ifmmode r_{\rm f}  \else $r_{\rm f}$ \fi}

\newcommand {\muI} {\ifmmode \mu_{0,I} \else $\mu_{0,I}$ \fi}

\newcommand {\mgal}    {\ifmmode m_{\rm gal}    \else $m_{\rm gal}$ \fi} 
\newcommand {\mgalo}    {\ifmmode m_{\rm gal,0}    \else $m_{\rm gal,0}$ \fi} 
\newcommand {\md}    {\ifmmode m_{\rm d}    \else $m_{\rm d}$ \fi} 
\newcommand {\Md}    {\ifmmode M_{\rm d}    \else $M_{\rm d}$ \fi} 
\newcommand {\Ms}    {\ifmmode M_{*}    \else $M_{*}$ \fi} 
\newcommand {\Mb}    {\ifmmode M_{\rm b}    \else $M_{\rm b}$ \fi} 
\newcommand {\Mf}    {\ifmmode M_{\rm f}    \else $M_{\rm f}$ \fi} 
\newcommand {\Mi}    {\ifmmode M_{\rm i}    \else $M_{\rm i}$ \fi} 
\newcommand {\Mgal}  {\ifmmode M_{\rm gal}  \else $M_{\rm gal}$ \fi}

\newcommand {\Jd} {\ifmmode J_{\rm d} \else $J_{\rm d}$ \fi} 
\newcommand {\Jb} {\ifmmode J_{\rm b} \else $J_{\rm b}$ \fi}  
\newcommand {\Jgal} {\ifmmode J_{\rm gal} \else $J_{\rm gal}$ \fi}  
\newcommand {\fc} {\ifmmode f_{\rm c} \else $f_{\rm c}$ \fi} 
\newcommand {\fx} {\ifmmode f_{\rm x} \else $f_{\rm x}$ \fi} 
\newcommand {\ft} {\ifmmode f_{\rm t} \else $f_{\rm t}$ \fi} 
\newcommand {\fV} {\ifmmode f_{\rm V} \else $f_{\rm V}$ \fi}
\newcommand {\jb} {\ifmmode j_{\rm b} \else $j_{\rm b}$ \fi} 
\newcommand {\jgal} {\ifmmode j_{\rm gal} \else $j_{\rm gal}$ \fi} 
\newcommand {\flost} {\ifmmode f_{\rm lost} \else $f_{\rm lost}$ \fi} 

\newcommand {\Vd}    {\ifmmode  V_{\rm d}    \else $V_{\rm d}$    \fi} 
\newcommand {\Vb}    {\ifmmode  V_{\rm b}    \else $V_{\rm b}$    \fi} 
\newcommand {\VDM}   {\ifmmode  V_{\rm DM}   \else $V_{\rm DM}$    \fi} 
\newcommand {\Vcirc} {\ifmmode  V_{\rm circ} \else $V_{\rm circ}$ \fi} 
\newcommand {\VII}   {\ifmmode  V_{2.2}      \else $V_{2.2}$      \fi}

\newcommand {\vdt} {\ifmmode  \Vd/\VII \else  $\Vd/\VII$ \fi}  
\newcommand {\dvr} {\ifmmode \partial \log  V_{2.2} / \partial \log R_{\rm d} 
     \else  $\partial \log V_{2.2} / \partial \log R_{\rm d}$ \fi} 

\newcommand {\YI} {\ifmmode  \Upsilon_I  \else $\Upsilon_I$ \fi} 
\newcommand {\LI} {\ifmmode L_I \else $L_I$ \fi}
\newcommand {\DeltaIMF} {\ifmmode \Delta_{\rm IMF} \else $\Delta_{\rm IMF}$ \fi}

\slugcomment{T{\tiny HE} A{\tiny STROPHYSICAL} J{\tiny OURNAL} {\scriptsize 654:27-52, 2007 January 1}}

\shorttitle{A Revised Model for Disk Galaxy Formation} 
\shortauthors{Dutton et al. 2006}

\begin{document}


\title{A Revised Model for the Formation of Disk Galaxies: \\
Low Spin and Dark-Halo Expansion}

\author{Aaron A. Dutton}  
\affil{Department of Physics, Swiss Federal Institute of Technology 
      (ETH Zurich), CH-8093 Zurich, Switzerland}

\author{Frank C. van den Bosch}  
\affil{Max-Planck-Institut  f\"ur Astronomie, K\"onigstuhl 17, 69117 
       Heidelberg, Germany}

\author{Avishai Dekel}  
\affil{Racah Institute of Physics, The Hebrew University, Jerusalem, Israel}

\and

\author{St\'ephane Courteau}    
\affil{Department of Physics, Eng. Physics \& Astronomy, Queen's
  University, Kingston, ON K7L 3N6, Canada}


\begin{abstract}
  We   use   observed    rotation   velocity-luminosity   ($VL$)   and
  size-luminosity ($RL$)  relations to single out  a specific scenario
  for  disk  galaxy  formation  in  the \LCDM  cosmology.   Our  model
  involves  four independent  log-normal  random variables:  dark-halo
  concentration $c$, disk spin  $\lamgal$, disk mass fraction $\mgal$,
  and stellar  mass-to-light ratio \YI.   A simultaneous match  of the
  $VL$  and  $RL$  zero  points with  adiabatic  contraction  requires
  low-$c$ halos,  but this model  has $V_{2.2} \sim 1.8  \Vvir$ (where
  $V_{2.2}$ and  $\Vvir$ are the  circular velocity at 2.2  disk scale
  lengths and the virial radius, respectively) which will be unable to
  match  the  luminosity  function  (LF).   Similarly  models  without
  adiabatic contraction  but standard $c$ also predict  high values of
  $V_{2.2}/\Vvir$.   Models in  which disk  formation induces  an {\it
  expansion}  rather  than the  commonly  assumed  contraction of  the
  dark-matter  halos  have  $V_{2.2}\sim  1.2 \Vvir$  which  allows  a
  simultaneous  fit of the  LF.  This  may result  from non-spherical,
  clumpy gas accretion, where dynamical friction transfers energy from
  the gas to  the dark matter.  This model  requires low $\lamgal$ and
  $\mgal$ values,  contrary to  naive expectations.  However,  the low
  $\lamgal$  is   consistent  with  the  notion   that  disk  galaxies
  predominantly survive in halos with  a quiet merger history, while a
  low $\mgal$ is also indicated by galaxy-galaxy lensing.  The smaller
  than  expected  scatter  in  the  $RL$ relation,  and  the  lack  of
  correlation between  the residuals of  the $VL$ and  $RL$ relations,
  respectively, imply that the scatter in $\lamgal$ and in $c$ need to
  be smaller than predicted for \LCDM halos, again consistent with the
  idea that disk galaxies preferentially  reside in halos with a quiet
  merger history.
  
\end{abstract}



\section{Introduction}
\label{sec:intro}

In  the  standard,  cold  dark  matter  (CDM)  based  model  for  disk
formation, set out by Fall \& Efstathiou (1980), disks form out of gas
that slowly cools out of a  hot gaseous halo, associated with the dark
matter  potential   well,  while  maintaining   its  specific  angular
momentum.   During this  process  the dark  matter  halo contracts  to
conserve its adiabatic invariants (Blumenthal \etal 1986).  Because of
the centrifugal  barrier the gas  settles in a  rotationally supported
disk whose size is proportional  to both the size and angular momentum
of  the dark  matter halo  (Mo, Mao  \& White,  1998; hereafter  MMW). 
Consequently, the structure and dynamics of disk galaxies are expected
to be  strongly correlated  with the properties  of their  dark matter
halos.   In  particular,  the  correlations  between  the  observable,
structural parameters of disk  galaxies, rotation velocity, $V$, size,
$R$,  and luminosity,  $L$, are  expected to  be a  reflection  of the
virial   properties   of   dark   matter   halos,   which   scale   as
$\Vvir\propto\Rvir\propto\Mvir^{1/3}$.   Slight deviations  from these
scalings are  expected from the fact  that more massive  halos are, on
average,  less  concentrated.   Any  further  deviations  must  either
reflect  some  aspects  of  the  baryonic physics  related  to  galaxy
formation, or signal a failure in the standard picture outlined above.
In  what follows we  refer to  the relations  between the  global disk
parameters as the $VL$, $RL$, and $RV$ relations.

In  the  past, the  $VL$  relation,  also  known as  the  Tully-Fisher
relation  (Tully \&  Fisher 1977),  has received  much attention  as a
distance  indicator owing to  the relatively  small observed  scatter. 
Although  numerous  studies have  addressed  the  origin  of the  $VL$
relation,  no  consensus  has  been  reached.  In  particular,  it  is
currently still under  debate whether the origin of  the $VL$ relation
is   mainly  governed  by   initial  cosmological   conditions  (e.g.,
Eisenstein \& Loeb 1996; Avila-Reese, Firmani \& Hern\'andez 1998), or
by the detailed processes governing star formation (Silk 1997; Heavens
\& Jiminez 1999) and/or feedback (e.g., Kauffmann, White \& Guiderdoni
1993; Cole \etal  1994; Elizondo \etal 1999; Natarajan  1999).  In the
most recent  models (van  den Bosch 2000;  2002; Navarro  \& Steinmetz
2000;  Firmani \& Avila-Reese  2000, hereafter  FA00) it  is typically
understood that  both initial conditions and  baryonic physics related
to  star  formation  and  feedback   must  play  an  important  role.  
Reproducing the $VL$ zero point  has also been a long standing problem
for CDM  based galaxy  formation models.  In  particular no  model has
been  able to simultaneously  match the  luminosity function  and $VL$
zero point  using standard \LCDM  parameters (Cole \etal  2000; Benson
\etal 2003;  Yang, Mo, \&  van den Bosch  2003).  This problem  can be
traced to the  high values of $V/\Vvir$ expected  for \LCDM halos once
the effects of the  baryons, such as adiabatic contraction (Blumenthal
\etal 1986),  are taken  into account. All  solutions to  this problem
require  a change  to either  the standard  cosmological model  or the
standard picture of galaxy formation.

Another potential  problem for disk formation models  is the formation
of large  enough disks.  The  relative fragility of disk  galaxies and
the  strongly ordered  motion  of  their stars  and  gas is  generally
interpreted  as evidence  for  a relatively  smooth formation  history
without violent merger processes.  The sub-sample of dark matter halos
without recent major mergers is  known to have systematically low spin
parameters (D'Onghia  \& Burkert 2004).  However,  standard models for
the formation  of extended  rotating disks seem  to require  high spin
parameters in order to reproduce the zero point of the $RL$ relation.

In  addition to  the  slopes and  zero  points of  the  $VL$ and  $RL$
relations, additional  constraints come from the scatter  in these two
relations. In particular, Courteau  \& Rix (1999; hereafter CR99) have
shown  that the  residuals of  the $VL$  relation, at  fixed  $L$, are
virtually  uncorrelated with the  residuals of  the $RL$  relation, at
fixed $L$ (see also McGaugh  2005).  This is another way of expressing
the fact that  the $VL$ relation is independent  of surface brightness
(see Courteau \etal 2006 for details),  and implies that the size of a
disk at given $L$ has no  relevance to its rotation velocity.  This is
a puzzling result to explain, as  one would naively expect that a more
concentrated disk  also results in a higher  rotation velocity.  CR99,
therefore,  interpreted the  weak residual  correlation  as indicating
that,  {\it  on average},  high  surface  brightness  (HSB) disks  are
sub-maximal, so that the disk  only contributes mildly to the observed
rotation velocity.  If confirmed  this puts constraints on the stellar
mass-to-light ratio, $\Upsilon$, and  thus on the stellar initial mass
function (IMF).  FA00, on the other hand, claim that the weak residual
correlation owes to the  surface density dependence of star formation,
such that  at a  given baryonic mass,  lower surface  density galaxies
have  lower  stellar  masses,  and  hence  lower  luminosities,  which
compensate for  the somewhat  lower rotation velocities.   However, in
their model  both the highest  and lowest surface  brightness galaxies
lie above the mean $VL$  relation, contrary to observations, and their
models do  not reproduce the  zero point of  the $VL$ relation  or the
amount of scatter in the $VL$ and $RL$ relations.

Although various studies  have attempted to explain the  origin of the
$VL$  or the  $RL$  relation, the  true  challenge lies  in finding  a
self-consistent model of disk  formation that can match both relations
as  well  as the  galaxy  luminosity  function  {\it simultaneously}.  
Finding such a model is a non-trivial task, as all current models fail
to do so.  In this paper we examine the parameter space of such models
within  the  standard \LCDM  cosmology.  We  simultaneously match  the
slopes, zero points and residuals  of the $VL$ and $RL$ relations.  We
also  investigate what  each  of  these models  predict  for the  mean
$V/\Vvir$. In order to be able to reproduce the observed abundances of
disk galaxies, this  ratio needs to be relatively  low $\sim 1.2$.  We
show  that this  restriction severely  limits the  allowable parameter
space, favoring a model with  low spin parameter $\lamgal$, low galaxy
mass  fraction  $\mgal$, and  with  halo  expansion  rather than  halo
contraction.

The key ingredients of our model are as follows:
\begin{enumerate}
\item  Disk  galaxies  form   in  spherical  dark-matter  halos  whose
  properties are  drawn from N-body simulations of  the standard \LCDM
  cosmology.  In  particular, their density profiles have  an NFW form
  with a concentration parameter $c$ that declines systematically with
  mass.
\item The baryons form an exponential disk in centrifugal equilibrium,
  which is specified  by a mass fraction $\mgal$  and a spin parameter
  $\lamgal$.
\item The galaxy  mass fraction is treated as a  free parameter with a
  mean $\mgal\propto\Mvir^{\alpham}$. Positive values of $\alpham$ are
  expected from feedback effects.
\item A  bulge is included  based on a self-regulating  mechanism that
  ensures disk stability.
\item The interaction between the  baryons and the dark matter halo is
  modeled with  a generalized  adiabatic contraction model  which also
  allows for halo expansion.
\item Stars form in the disk once above a threshold surface density.
\item  The I-band  stellar  mass-to-light ratio  $\YI$ increases  with
  luminosity as constrained by observations.
\item  The model  parameters  $c$, $\lamgal$,  $\mgal$  and $\YI$  are
  assumed to be independent log-normal random variables.
\end{enumerate}

The   data  and   models  are   introduced  in   \S\ref{sec:data}  and
\S\ref{sec:models},  respectively.  In  \S\ref{sec:ml} we  outline the
conversion    between     stellar    mass    and     luminosity;    In
\S\ref{sec:sampling} we  detail the  computation of our  model scaling
relations.  In \S\ref{sec:convert} we  discuss how to construct models
that match the slopes,  zero points, scatter and residual correlation.
In  \S\ref{sec:lowspin}  we explore  the  model  parameter space,  and
advocate a revised model for disk formation.  We summarize our results
in \S\ref{sec:summary}.


\section{The Data}
\label{sec:data}

We compare  our models with  the large data  set of $\sim  1600$ local
disk galaxies  compiled by Courteau  \etal (2006).  For each  of these
galaxies  a rotation  velocity $V$,  a  luminosity $L$,  a disk  scale
length $R$, and a central  surface brightness of the disk $\mu_0$, are
available.  The  data set is compiled from  three independent samples:
Mathewson,  Ford, \&  Buchhorn 1992  (hereafter, MAT);  Courteau \etal
2000 (hereafter, Shellflow); and Dale \etal 1999 (hereafter, SCII).

All galaxy  luminosities are measured in the  $I$-band, and correspond
to {\it total} luminosities (i.e.,  disk plus bulge). In addition, for
a significant sub-sample we  also have optical colors available, which
we use  to estimate stellar mass-to-light  ratios (see \S\ref{sec:ml}
below).  Disk  scale lengths  and surface brightnesses  are determined
from the $I$-band  photometry.  In what follows, whenever  we refer to
surface  brightness  we mean  the  {\it  central} surface  brightness,
$\muI$, of the exponential disk fitted to the data.

Due to the complications  involved with interpreting \hi\ line widths,
we  only use rotation  velocities derived  from resolved  \ha rotation
curves. This reduces the full  sample by $\sim 300$ galaxies.  For the
MAT  and Shellflow  samples  the  rotation curves  are  fitted with  a
parametric  function (Courteau 1997)  which is  then evaluated  at 2.2
disk scale lengths. For the SCII sample the velocities are measured at
the optical radius (equivalent to 3.2 scale lengths for an exponential
disk). 

\begin{figure*}
\begin{center}
\includegraphics[width=6.0in]{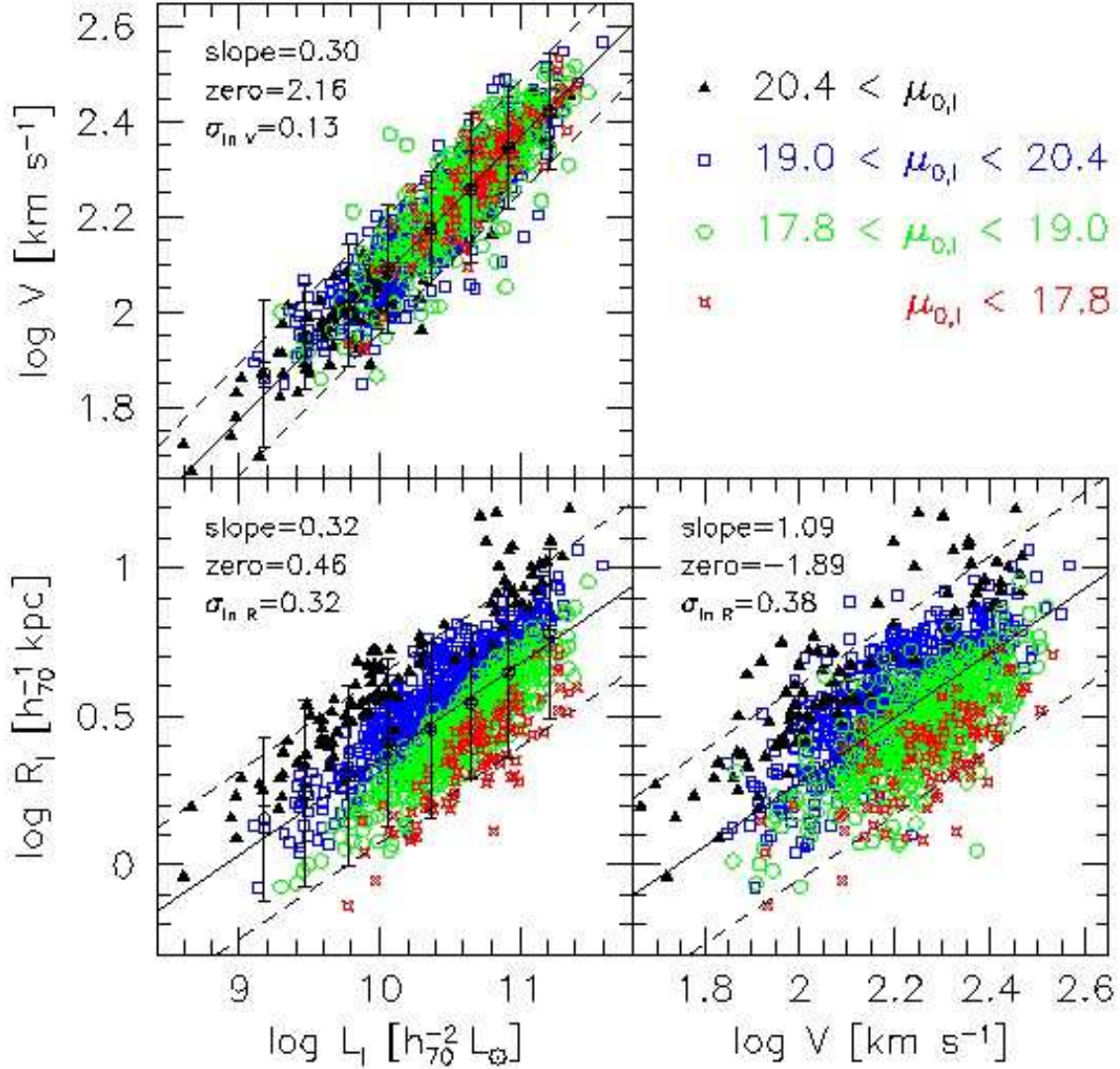}
\caption{Observed I-band $VLR$ scaling relations using data from 
  Courteau \etal  2006. Bi-weighted orthogonal least  squares fits are
  given by the solid black lines, with 2-sigma deviations given by the
  dashed lines. The  open black circles with error  bars show the mean
  and 2-sigma scatter of the $VL$ and $RL$ relations binned at 0.3 dex
  intervals  in  $L_I$.  The  colors  and  point  types correspond  to
  extrapolated disk central surface brightness, $\muI$ as indicated in
  the top right panel.  }
\label{fig:VLR_ALL}
\end{center}
\end{figure*}

\begin{figure*}
\begin{center}
\includegraphics[width=6.0in]{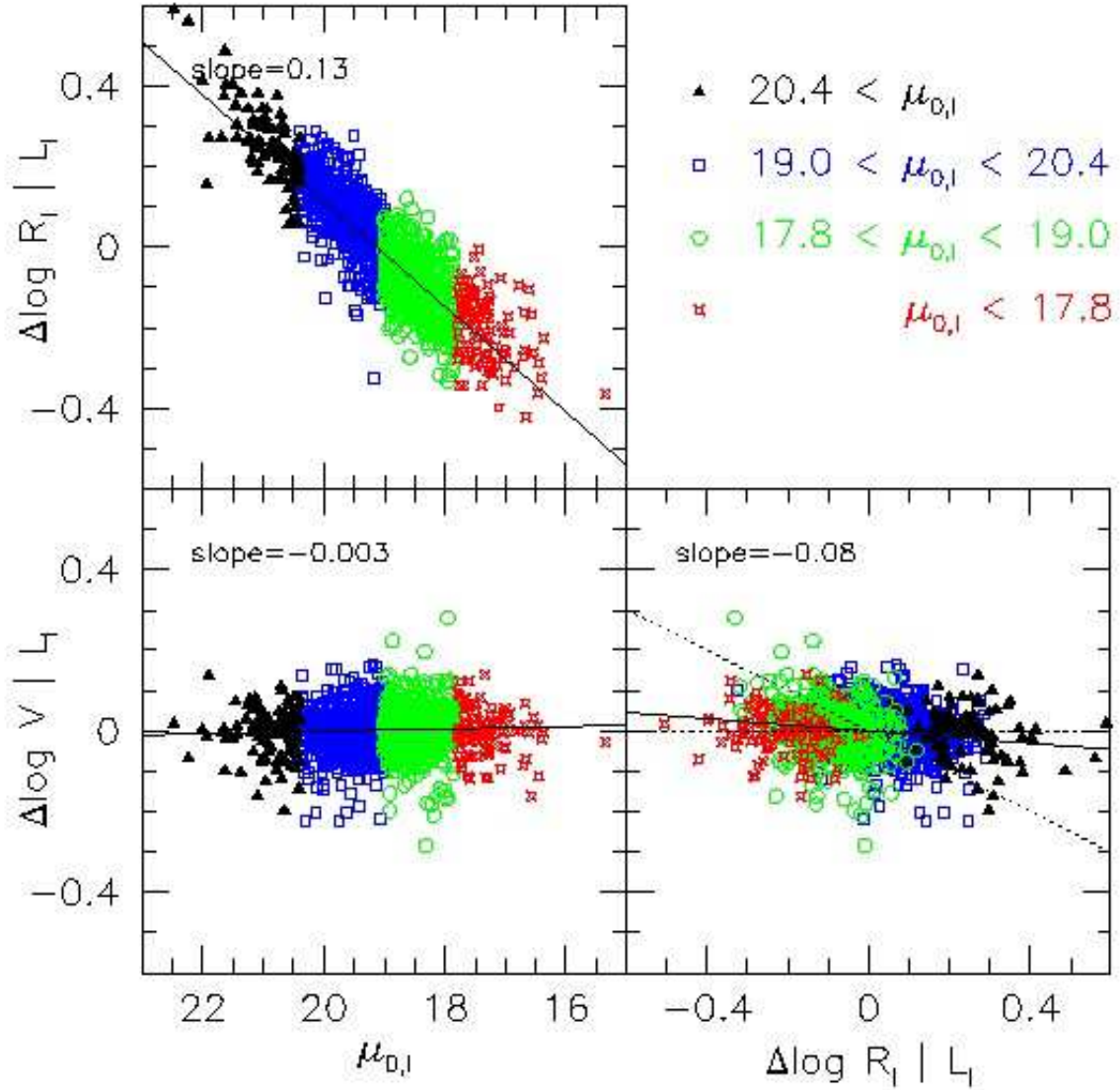}
\caption{Residual correlations from the observed I-band $VLR$
  relations in  Fig.~\ref{fig:VLR_ALL}. The black  lines show weighted
  least-squares  fits.  The  $RL$ residuals  show a  clear correlation
  with surface  brightness, while the  $VL$ residuals show  none.  The
  residuals of the $VL$ and  $RL$ relation are only weakly correlated. 
  The dotted line has a slope  of -0.5 expected for a pure exponential
  disk (CR99).}
\label{fig:dVLR_ALL}
\end{center}
\end{figure*}

\begin{figure*}
\begin{center}
\includegraphics[width=6.0in]{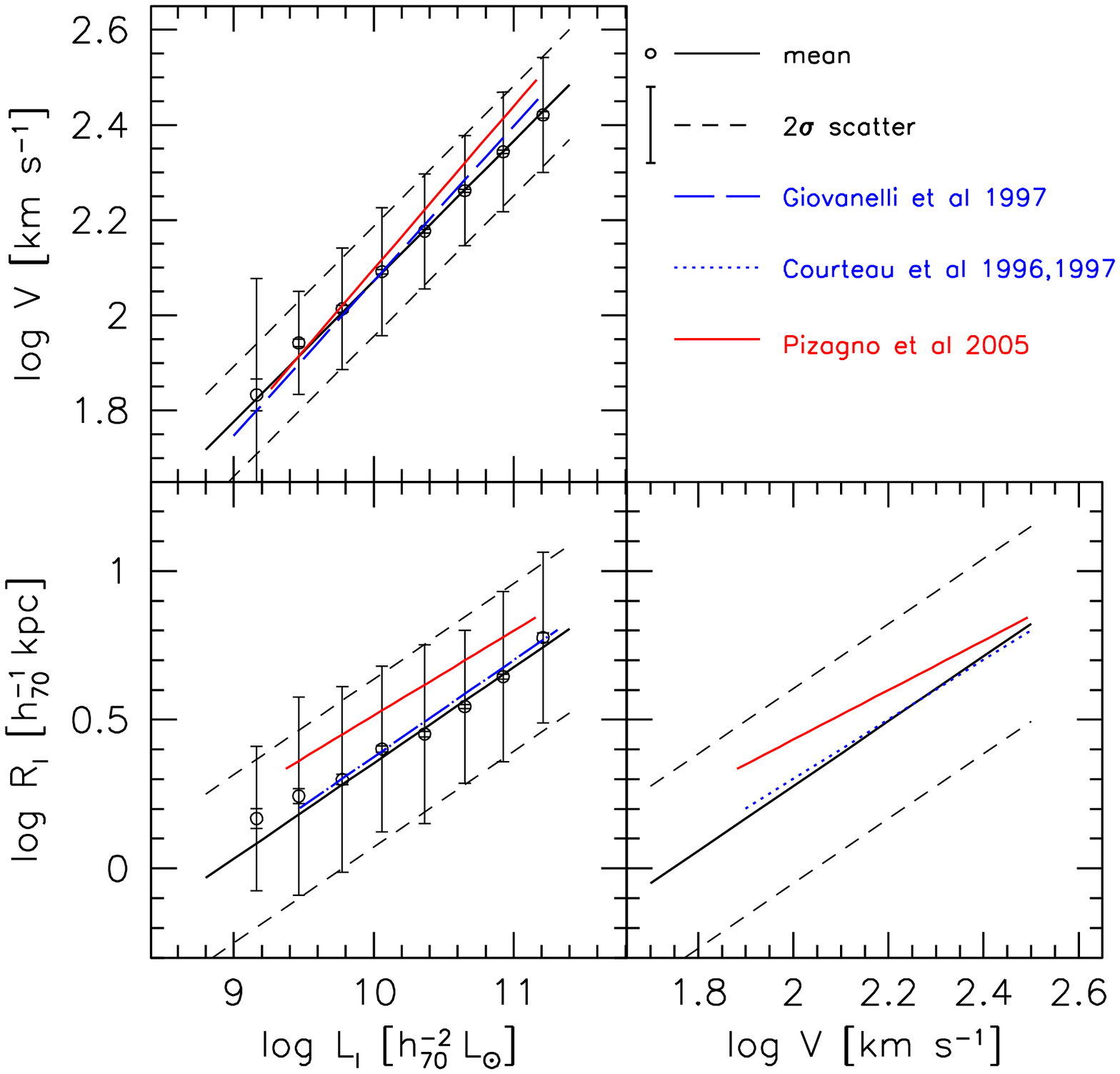}
\caption{Comparison of  our I-band $VLR$ scaling  relations with those
  from the  literature.  The long-dashed lines show  the $VL$ relation
  from Giovanelli \etal 1997 (G97), and the dotted lines show the $RV$
  relation from Courteau (1996, 1997).  The red lines are derived from
  Pizagno \etal (2005).}
\label{fig:VLR_ALL_lit}
\end{center}
\end{figure*}

\subsection{Corrections}
\label{sec:datacorr}

In order to homogenize these data samples as much as possible, we have derived
the inclination  and cosmological corrections to  the velocities, luminosities
and scale lengths in a uniform way, as described below.

The observed rotation velocities, $V_{\rm obs}$, are corrected for inclination
and cosmological broadening using
\begin{equation}
V = \frac{V_{\rm obs}} {(1+z)\, \sin i }. 
\end{equation}
The inclination, $i$, is computed using
\begin{equation}
\sin i = \sqrt{\frac{1- (b/a)^2}{1-q_0^2}},
\end{equation}
where $b/a$ is the minor-to-major axis ratio, and $q_0$ is the intrinsic
thickness of the disk. We assume $q_0=0.2$, and set $i=90^{\circ}$ if
$b/a<q_0$.

The  apparent  magnitudes,  $m_I$,  are corrected  for  both  internal
extinction, $A_{\rm  int}$, and external  (i.e., Galactic) extinction,
$A_{\rm ext}$. A small k-correction, $A_{\rm k}$, is also applied such
that
\begin{equation}
m_I = m_{I,{\rm obs}} - A_{\rm int} - A_{\rm ext} - A_{\rm k}.
\end{equation}
Internal  extinctions  are  computed  using  the  line-width  ($W=2V$)
dependent relation from Tully \etal (1998):
\begin{equation}
A_{\rm  int}   =  \gamma_I(W)  \log(a/b)  
               =  \left[0.92   +  1.63(\log  W -2.5)\right] \log(a/b)
\end{equation}
with  $a/b$ the  major-to-minor axis  ratio.  The  external (Galactic)
extinction is computed using the dust maps of Schlegel, Finkbeiner, \&
Davis  (1998), while  the k-corrections  are computed  using  the line
width dependent formalism of Willick \etal (1997).

The absolute magnitudes, $M_I$, are computed using
\begin{equation}
  M_I   =   m_I   -   5   \log  D_{\rm L}   -25;   \;\;\;\;   D_{\rm L}=\frac{V_{\rm
      CMB}}{100h}(1+z),
\end{equation}
with  $V_{\rm  CMB}$  the  systemic  velocity of  the  galaxy  in  the
reference frame  at rest with  the cosmic microwave  background (Kogut
\etal 1993).  The $I$-band  luminosities are computed from $M_I$ using
an absolute magnitude for the Sun of $M_{I,\odot}=4.19$.

Disk scale lengths are corrected for inclination using
\begin{equation}
R_{\rm I} = R_{\rm I,obs} / [1+0.4\log(a/b)],
\end{equation}
(Giovanelli  \etal 1994),  and  converted into  kilo parsecs using  the
angular diameter distance $D_{\rm  A} = V_{\rm CMB}/[100 \, h \, (1+z)]$.

Finally, central  surface brightnesses are  corrected for inclination,
Galactic extinction, and cosmological dimming ({\it per unit frequency
interval}) using
\begin{equation}
\mu_{0,I} = \mu_{0,I,\rm obs} + 0.5 \log(a/b) - A_{\rm ext} - 2.5\log(1+z)^3.
\end{equation}
The factor  of $0.5$ in front  of the $\log(a/b)$  term is empirically
determined  by demanding that  the residuals  of the  relation between
central surface  brightness and  rotation velocity has  no inclination
dependence\footnote{For  a  disk of  zero  thickness  one expects  the
  factor to be between 0, for an optically thick disk, and 2.5, for an
  optically thin disk.}.  Following  Giovanelli \etal (1997) we assume
an  uncertainty  of 15\%  in  $\gamma_I$,  $a/b$,  $A_{\rm ext}$,  and
$A_{\rm  k}$,  and  propagate  the  errors.   Not  including  distance
uncertainties, the average errors  on the observables are $\sigma_{\ln
  V}  \simeq  0.08$, $\sigma_{\ln  L}  \simeq  0.1$, and  $\sigma_{\ln
  R}\simeq0.14$.

\subsection{Scaling Relations}
\label{sec:datafit}

We now use  the data described above to investigate  the $VL$ and $RL$
relations of disk galaxies.  Despite the fact that our sample consists
of subsamples  that use  slightly different techniques  or definitions
for the scale lengths and  rotation velocities, we find that, with the
uniform inclination  and extinction corrections  described above, each
subsample yields $VL$ and $RL$ relations that are consistent with each
other (see Courteau \etal 2006 for details).  We therefore combine the
three subsamples to a single sample of $\sim 1300$ galaxies.

The resulting $VLR$ relations  are shown in Fig.\ref{fig:VLR_ALL}. The
solid lines show the mean  relations, which have been determined using
bi-weighted orthogonal  least-squares fits. The dashed  lines show the
$2\sigma$ scatter. The open circles with error bars show the data mean
and $2  \sigma$ scatter  in separate luminosity  bins with a  width of
$0.3$  dex.  These  show that  the $VL$  and $RL$  relations  are well
fitted with  a single power-law  over the range considered  here.  The
$RV$  relation has the  largest scatter,  and is  thus the  least well
determined.  For consistency  we use the mean $VL$  and $RL$ relations
to determine the mean $RV$ relation.

The mean relations thus derived are:
\begin{equation}
\log          \frac{V}{[\kms]}          =          0.296          \log
\frac{L_I}{[10^{10.3}h_{70}^{-2}L_{\odot}]} +2.162,
\end{equation}
\begin{equation}
\log       \frac{\RI}{[h_{70}^{-1}\kpc]}       =      0.322       \log
\frac{L_I}{[10^{10.3}h_{70}^{-2}L_{\odot}]} +0.455.
\end{equation}
\begin{equation}
\log  \frac{\RI}{[h_{70}^{-1}\kpc]}   =  1.086  \log  \frac{V}{[\kms]}
\hspace{0.7truecm}-1.894.
\end{equation}
where $\RI$ is  the de-projected disk scale length  in the photometric
$I$-band, and $h_{70} = H_0 / (70 \kmsmpc)$.

The color coding in  Fig.\ref{fig:VLR_ALL} denotes the central surface
brightness of the disk in  the $I$-band, $\muI$. Note that the scatter
in  the $VL$ relation  is not  correlated with  $\muI$.  This  is also
apparent  from Fig.~\ref{fig:dVLR_ALL}  which  shows the  correlations
between the residuals  of the $VL$ and $RL$  relations at constant $L$
($\Delta  \log  V$ and  $\Delta  \log  R$,  respectively) and  surface
brightness, $\muI$.   Again the symbols  are color coded  according to
$\muI$.  As is  evident from the upper left panel,  the scatter in the
$RL$ relation  is by definition\footnote{A family  of pure exponential
  disks with constant  $\YI$ has, at a constant  $\LI$, $\Delta \log R
  \propto 0.2  \muI$, which  is different from  the observed  slope of
  $0.13$.   This owes to  the relatively  small range  of luminosities
  sampled, relative  to the amount of  scatter in the  $RL$ relation.} 
dominated by scatter in surface brightness.  The residuals of the $VL$
relation,  however,  show  no  significant  correlation  with  surface
brightness at all  (lower left panel). In addition,  the $VL$ and $RL$
residuals  are only very  weakly correlated  (lower right  panel): the
slope of  the residual correlation  is $\gamma \equiv  \dd[\Delta \log
V(L)] /  \dd[\Delta \log  R(L)] = -0.08  \pm 0.03$, in  agreement with
CR99.

Fig.~\ref{fig:VLR_ALL_lit} shows  a comparison of  the $VLR$ relations
derived here with  previous studies.  The black lines  and point types
are the same as in Fig.~\ref{fig:VLR_ALL}.  The long-dashed and dotted
lines show  the $VL$  relation from Giovanelli  \etal (1997),  and the
$RV$ relation that  fits the data of Courteau  (1996, 1997). Note that
these $VL$ and  $RL$ relations, which are in  excellent agreement with
our  mean  scaling relations,  were  used by  MMW  and  FA00 as  model
constraints.   The red  lines in  Fig.~\ref{fig:VLR_ALL_lit}  show the
$VLR$ relations of Pizagno \etal (2005).  The latter study is based on
a sample of 81 disk dominated (B/D < 0.11) galaxies, with luminosities
in the Sloan $i$-band,  disk scale lengths measured with bulge-to-disk
decompositions, and velocities measured at 2.2 disk scale lengths from
\ha rotation curves. We convert from  $L_i$ to $L_I$ using $\log L_I =
\log   L_i  +0.4(i-I)   -0.4(i-I)_{\odot}   \simeq  0.036$,   assuming
$M_{i,\odot}=4.56$, and $(i-I)\simeq0.46$ (Courteau \etal 2006).  Note
that  the  slope of  the  $VL$ relation  of  Pizagno  \etal (2005)  is
somewhat steeper than  ours, while their $RL$ relation  is offset from
ours towards larger disk scale  lengths.  These differences may be due
to a combination of their  relatively small sample size (81 galaxies),
their   bulge-to-disk   ratio   selection  criteria,   and   different
inclination corrections.  These  differences do not significant impact
on our main conclusions, as we discuss in Appendix~\ref{sec:Piz05}


\section{Disk Formation Models}
\label{sec:models}

In the  standard picture of  disk formation (e.g., Fall  \& Efstathiou
1980),  disks form  inside virialized  dark matter  halos  through the
cooling  of the  baryonic  material.   Our models  are  based on  this
standard  picture, and closely  follow MMW,  but with  some additional
ingredients. In  particular, we include  a model for the  formation of
bulges, we use  a prescription for star formation  which separates the
disk  into gaseous  and stellar  components,  we use  an improved  and
generalized  description   for  adiabatic  contraction,   and  we  use
empirical relations to convert our models to observable quantities for
direct  comparison with  the data  described above.   All  these model
ingredients are discussed in more detail below.

Although we  refer to our models  as `disk formation  models' they are
completely `static' (i.e., the model  is not actually evolved).  For a
given specific angular momentum of  the baryonic material out of which
the  disk forms  and a  given potential  due to  the dark  matter, the
structural properties of the  resulting disk are computed assuming (i)
that  disks  are  exponential,  and  (ii) that  the  specific  angular
momentum of  the baryonic  material is conserved.   Alternatively, one
could  in principle consider  more `dynamic'  models, that  follow the
actual  formation of  the  disk galaxies  starting  at high  redshifts
(e.g., Firmani  \& Avila-Reese 2000; Avila-Reese \&  Firmani 2000; van
den Bosch 2001,  2002). However, as long as the  formation of the disk
is sufficiently quiescent,  the final structure of the  disk should be
independent of its actual formation history: the structural properties
of  the final disk  are basically  just governed  by the  principle of
dynamical, centrifugal  equilibrium.  In a more  dynamic approach, one
can actually  model the star formation  history of the  disk, which is
not  possible with  our  static model.   However,  the star  formation
history  mainly governs the  final mass-to-light  ratio of  the stars,
which  we set using  empirical relations.   The models  described here
should  thus  be  applicable  independent of  the  detailed  formation
history as long as the two assumptions mentioned above are satisfied.

Unless stated  otherwise we  adopt a \LCDM  cosmology with  $\OmegaM =
0.3$, $\Omega_{b}=0.044$, $\OmegaL =0.7$, $h = H_0/(100\kmsmpc) = 0.7$
and with a scale-invariant initial power spectrum with a normalization
$\sigma_8  =  0.9$.  The  baryonic  mass  fraction  of this  cosmology
$\fbar\simeq0.15$.

\subsection{Disk Formation}
\label{sec:diskform}

We model dark matter halos as spheres with a NFW density distribution
\begin{equation}
\label{nfw}
\rho(r) = {4 \rho_s \over (r/\rs) (1 + r/\rs)^{2}}
\end{equation}
(Navarro, Frenk \& White 1997)  where $\rs$ is a characteristic radius
at  which  the logarithmic  slope  of  the  density distribution  $\dd
\ln\rho/\dd \ln r = -2$,  and $\rho_{\rm s} = \rho(\rs)$.  The overall
shape  of  the  density  distribution  can  be  characterized  by  the
so-called concentration parameter  $c=\Rvir/\rs$.  Here $\Rvir$ is the
virial  radius, which is  defined as  the radius  inside of  which the
average  halo density is  $\Deltavir$ times  the critical  density for
closure.  For  the \LCDM cosmology adopted  here $\Deltavir\simeq 100$
(Bryan \& Norman  1998).  In addition to the  virial radius $\Rvir$ we
also define  the virial velocity  $\Vvir$ as the circular  velocity at
the virial radius, i.e., $\Vvir^2 =  G \, \Mvir / \Rvir$, with $G$ the
gravitational constant.

The total angular  momentum of a halo, $\Jvir$,  is commonly expressed
in terms of the dimensionless spin parameter:
\begin{equation}
\label{spin}
\lambda = {J_{\rm vir}  \vert E \vert^{1/2} \over  G M_{\rm
vir}^{5/2}} = \frac{\Jvir/\Mvir}{\sqrt{2}\, \Rvir \Vvir} \fc^{1/2}
\end{equation}
Here $E$ is the halo's energy, and $\fc$ measures the deviation of $E$
from that of  a singular isothermal sphere with the  same mass, and is
given by
\begin{equation}
\label{fc}
\fc  = {c  \over  2} {1  -  1/(1+c)^2 -  2  {\rm ln}(1+c)/(1+c)  \over
\left[c/(1+c) - {\rm ln}(1+c) \right]^2}
\end{equation}
(see MMW).

We assume that  the galaxy that forms consists of a  bulge and a disk.
Our algorithm to ascribe a bulge-to-disk ratio is discussed at the end
of this section. We define the baryonic mass of the total galaxy as
\begin{equation}
\label{mgal}
\Mgal \equiv \Md + \Mb = \mgal \Mvir
\end{equation}
with $\Md$ and  $\Mb$ the masses of the  disk and bulge, respectively,
and $0 < \mgal \lta \fbar$. We define the bulge-to-disk mass ratio as
$\Theta \equiv \Mb/\Md$ so that
\begin{equation}
\label{mdmd}
\Md = {1 \over 1 + \Theta} \mgal \Mvir
\end{equation}
\begin{equation}
\label{mbmb}
\Mb = {\Theta \over 1 + \Theta} \mgal \Mvir.
\end{equation}
In addition, we  write that the total angular  momentum of the baryons
out of which the disk plus bulge form is
\begin{equation}
\label{jgal}
\Jgal \equiv \jgal \Jvir.
\end{equation}
As shown by van den  Bosch \etal (2002), the specific angular momentum
distribution of  the total baryonic  mass (including those  baryons in
the halo that do not partake  in the formation of the disk plus bulge)
is virtually identical to that of the dark matter. Therefore, one also
expects that $0 < \jgal \lta \fbar$.

As  discussed  below, we  assume  that the  bulge  forms  out of  disk
instabilities.  Let  $\Jb = \jb \Jvir$ indicate  the original angular
momentum of  the baryonic  material out of  which the bulge  forms. We
assume,  however,  that the  bulge  formation  process transfers  this
angular momentum to the disk plus  the halo, so that the final angular
momentum  of the  bulge  is  zero. Indeed,  such  an angular  momentum
transfer  is  observed  in  numerical simulations  (e.g.,  Hohl  1971;
Debattista \etal  2006). If we define  $\ft$ as the  fraction of $\Jb$
that  is transferred to  the disk,  we obtain  that the  final angular
momentum of the disk is equal to
\begin{equation}
\label{jd}
\Jd = \left[ \jgal - (1 - \ft) \jb \right] \Jvir
\end{equation}
The {\it specific} angular momentum of the final disk is therefore
\begin{equation}
\label{jdmd}
{\Jd  \over \Md}  = (1  + \Theta)  [1 -  \flost] \left(  {\jgal \over
\mgal} \right) {\Jvir \over \Mvir}
\end{equation}
where we have introduced the parameter
\begin{equation}
\label{flost}
\flost = (1 - \ft) \left( {\jb \over \jgal} \right)
\end{equation}
which  expresses the  fraction of  the total  angular momentum  of the
material out of which the bulge  plus disk forms that has been lost to
the  halo. For  modeling  purposes it  is  more useful  to define  the
parameter
\begin{equation}
\label{eqn:fx}
\fx \equiv \flost  \left( {1+\Theta \over \Theta} \right)  = (1 - \ft)
\left( {\Jb/\Mb \over \Jgal/\Mgal} \right)
\end{equation}
which expresses the ratio of  the {\it specific} angular momentum that
has been  lost to the  halo due to  bulge formation to the  total {\it
  specific} angular  momentum of  the material out  of which  the disk
plus  bulge  have  formed.   For  example, $\fx=1$  means  that  bulge
formation does not  change the specific angular momentum  of the disk. 
The extreme  $\fx=0$ occurs when the  disk loses mass  but not angular
momentum  during  bulge  formation.  Note that  unlike  $\flost$,  the
parameter $\fx$  can in principle  be larger than unity.  In practice,
however, the bulge  is likely to form out  of material with relatively
low specific angular momentum  (e.g., Norman, Sellwood \& Hassan 1996;
van  den  Bosch \etal  2002).  Furthermore,  the  fraction of  angular
momentum  that is  lost to  the halo  is expected  to be  fairly small
(e.g.,  Weinberg  1985; Debattista  \&  Sellwood  2000; Valenzuela  \&
Klypin 2003; O'Neill  \& Dubinski 2003), so that  $\ft$ is expected to
be close to unity.

For a disk with a  surface density $\Sigma(R)$ and a circular velocity
$V(R)$ the total angular momentum is
\begin{equation}
\label{eqn:jofdisk}
\Jd = 2 \pi \int_{0}^{\Rvir} \Sigma(R) \, R \, V(R) \, R \dd R.
\end{equation}
Throughout  we assume  that the  disk  that forms  has an  exponential
surface density  distribution $\Sigma(R) =  \Sigma_0 \exp(-R/\Rd)$, so
that
\begin{equation}
\label{jdisk}
\Jd = 2 \Md \Rd \Vvir \fV
\end{equation}
with
\begin{equation}
\label{fr}
\fV  = {1  \over 2}  \int_{0}^{\Rvir/\Rd} {\rm  e}^{-u} u^2  {V(u \Rd)
\over \Vvir} \dd u
\end{equation}
(cf.,  MMW).   If  we  combine  equations~(\ref{spin}),  (\ref{jdmd}),
and~(\ref{jdisk})  we obtain  the following  expression for  the scale
length of the disk:
\begin{equation}
\label{rd}
\Rd  = {1 \over  \fV \sqrt{2  \fc}} \,  \left[ 1  + (1-\fx)  \, \Theta
\right] \, \lambda_{\rm gal} \, \Rvir
\end{equation}
where we have defined $\lamgal$ as the effective spin parameter of the
material out of which the bulge plus disk form:
\begin{equation}
\lamgal \equiv \left( {\jgal \over \mgal} \right) \lambda
\end{equation}
From equation~(\ref{rd})  it is  evident that bulge  formation impacts
the final scale length of the  disk. Typically, for $\fx > 1$ the disk
size will  decrease, while $\fx<1$  causes an increase in  $\Rd$. Note
that the  transition does  not occur exactly  at $\fx=1$,  because the
presence of  a bulge component modifies  $\fV$.  Numerical simulations
indicate that bulge formation causes an increase in disk scale lengths
(Debattista \etal 2006), suggesting that  $\fx < 1$. Shen \etal (2003)
considered a  fairly similar  model but with  the assumption  that the
material that forms  the bulge has the same  specific angular momentum
as the disk. Their favored model  is equivalent to our model with $\fx
= 0.5$.   However, given  the arguments above,  we expect $\fx$  to be
smaller  than this.  In  what follows  we  adopt a  fiducial value  of
$\fx=0.25$, although  none of our  results are very sensitive  to this
particular choice.

Note that since the computation  of $\fV$ requires knowledge of $\Rd$,
this  set of  equations  has to  be  solved iteratively  (see MMW  for
details).  When  computing the total  circular velocity $V(R)$  we use
that
\begin{equation}
V^2(R) = \Vd^2(R) + \Vb^2(R) + \VDM^2(R)
\end{equation}
Here
\begin{equation}
\Vd^2(R) = {G \Md \over \Rd} \, 2 y^2 \, \left[ I_0(y) K_0(y) - I_1(y)
K_1(y)\right]
\end{equation}
with $y=R/(2\Rd)$,  and $I_n$ and $K_n$ are  modified Bessel functions
(Freeman 1970). The  circular velocities of the bulge,  $\Vb$, and the
dark matter  halo, $\VDM$,  are computed assuming  spherical symmetry,
whereby the mass distribution of  the dark matter halo is adjusted for
adiabatic  contraction  (see   \S\ref{sec:ac}  below).   The  density
distribution  of the  bulge is  assumed to  follow a  Hernquist (1990)
profile
\begin{equation}
\rho_{\rm b} = \frac{\Mb}{2\pi} \frac{\rb}{r(r+\rb)^3}
\end{equation}
In projection,  this is similar to  a de Vaucouleurs  profile (i.e., a
Sersic profile  with Sersic  index $n=4$), with  a half  light radius,
$R_{\rm eff} =  1.8152 r_{\rm b}$ (Hernquist 1990). Throughout  we adopt the
relation between \Reff and \Mb from Shen \etal (2003):
\begin{equation}
\log R_{\rm  eff} = \left\{ \begin{array}{ll} 
-5.54 + 0.56 \log M_{\rm b} & \;\;(\log M_{\rm b} >  10.3) \\ 
-1.21 + 0.14 \log M_{\rm b} & \;\;(\log M_{\rm b} \le 10.3)
\end{array} \right. 
\end{equation}
Although  bulges of late-type  disk galaxies  are better  described by
exponential profiles  (e.g. Courteau, de  Jong \& Broeils  1996), what
matters  most for  our purposes  is the  {\it total}  bulge  mass; its
distribution is only of secondary importance.

The  computation of  the  actual bulge-to-disk  ratio  for each  model
galaxy is based  on the fact that self-gravitating  disks are unstable
against  global instabilities  (e.g., bar  formation).  We  follow the
approach  of van  den Bosch  (1998, 2000)  and Avila-Reese  \& Firmani
(2000) and assume  that an unstable  disk transforms part of  its disk
material  into a bulge  component in  a self-regulating  fashion, such
that  the final  disk is  marginally  stable.  
Bars, which are considered the  transitions objects in this scenario, are thus
expected to be fairly common, as observed.

We  define $\beta(R)  =
\Vd(R)/V(R)$, and consider the disk to be stable as long as
\begin{equation}
\label{eqn:betamax}
\beta_{\rm max}  = \max_{0  \leq R \leq  \Rvir} \beta(R)  < \beta_{\rm
crit}
\end{equation}
(Christodoulou,  Shlosman  \&  Tohline  1995).  The  actual  value  of
$\beta_{\rm crit}$ depends  on the gas mass fraction  of the disk, but
falls roughly  in the  range $0.52 \lta  \beta_{\rm crit} \lta  0.70$. 
For a given value of $\beta_{\rm crit}$, we use an iterative technique
to  find  the  bulge-to-disk   ratio  for  which  $\beta_{\rm  max}  =
\beta_{\rm crit}$.

\subsection{Adiabatic Contraction}
\label{sec:ac}

\begin{figure}
\begin{center}
\includegraphics[width=3.0in]{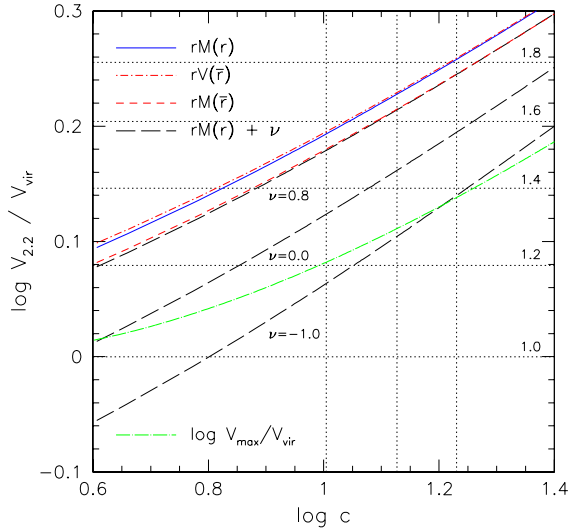}
\caption{Ratio of observed to virial circular velocities,
  $V_{2.2}/\Vvir$, as  a function of the  concentration parameter, for
  various   forms   of  adiabatic   contraction.    All  models   have
  $\lamgal=0.048$  and $\mgal=0.05$.  The  vertical dotted  lines show
  the mean $c$  according to model of Bullock  \etal (2001a) for halos
  with virial mass, $\Mvir =  10^{13}, 10^{12}$, \& $10^{11} \, h^{-1}
  M_{\odot}$.  The horizontal dotted lines show $V_{2.2}/\Vvir$ ratios
  of 1  to 1.8 at intervals  of 0.2.  The relation  using the standard
  (Blumenthal \etal  1986) adiabatic  invariant, $rM(r)$, is  given by
  the   blue   solid   line.    The  modified   adiabatic   invariant,
  $rM(\bar{r})$, as  proposed by Gnedin  \etal 2004 results in  only a
  $\simeq0.02$ dex  ($\simeq 5\%$)  reduction in $V_{2.2}$  (red short
  dashed  line).  Most  of  this reduction  is  taken back  if we  use
  specific angular momentum as  the adiabatic invariant, and take into
  account the  disk geometry  when computing $\Vcirc$  (dot-dashed red
  line).   For all  of  these adiabatic  contraction models,  standard
  concentration  parameters  yield  $V_{2.2}/\Vvir \simeq  1.6$.   The
  black  long-dashed  lines show  models  where  we have  artificially
  modified  the   contraction  factor  (see   equation~\ref{eqn:nu}).  
  Starting  from  the  standard  model  ($\nu=1$),  the  $rM(\bar{r})$
  relation is approximately reproduced with $\nu=0.8$, $\nu=0$ results
  in no adiabatic  contraction of the halo, while  $\nu=-1$ gives halo
  expansion.    The   green  dot-long   dashed   line  shows   $V_{\rm
    max}/\Vvir$,  where $V_{\rm  max}$  is the  maximum halo  circular
  velocity (without adiabatic contraction). }
\label{fig:Vvir}
\end{center}
\end{figure}

When baryons cool and concentrate in  the center of a dark matter halo
they deepen and  modify the shape of the  gravitational potential.  If
this process is slow with respect to the (local) dynamical time of the
halo,  the halo will  contract to  conserve its  adiabatic invariants.
For the  idealized case of a  spherical halo in which  all dark matter
particles move on circular  orbits, the adiabatic invariants reduce to
the  specific angular  momentum, $r  \, V(r)$.   If, in  addition, the
distribution  of  the baryons  has  spherical  symmetry, this  reduces
further to $r\,M(r)$, with $M(r)$  the enclosed mass within radius $r$
(Blumenthal \etal 1986; hereafter BFFP).

In realistic dark matter  halos, however, the particles typically move
on highly  eccentric orbits  (Ghigna \etal 1998;  van den  Bosch \etal
1999).   Taking this  into  account reduces  the  effect of  adiabatic
contraction  (Wilson  2003).   Gnedin  \etal (2004)  give  a  modified
adiabatic  invariant,  $r\,M(\bar{r})$,  where  $\bar{r}  \simeq  0.85
(r/\Rvir)^{0.8}$ is  the orbit averaged radius.   Using this adiabatic
invariant results in somewhat less contraction of the halo than in the
standard BFFP formalism.  An additional  problem is that disks are not
spherical.  Therefore, in principle one  should use $r \, V(r)$ as the
adiabatic invariant, rather  than $r \, M(r)$.  It  is well known that
the  circular velocity  curve of  a thin  exponential disk  rises less
rapidly but reaches a higher peak velocity than a sphere with the same
enclosed mass (e.g., Binney \& Tremaine 1987).  Therefore, using $r \,
V(r)$ as an adiabatic invariant, rather than $r \, M(r)$, results in a
stronger  contraction of  the dark  matter  halo at  $2.2$ disk  scale
lengths.

In addition  to these somewhat  subtle problems for the  standard BFFP
formalism, one may also  question whether adiabatic contraction really
occurs  during disk  formation.  If  disks  are not  built by  smooth,
relatively slow, spherical infall, the adiabatic contraction could, in
principle, be  counter-balanced by a variety of  processes.  These may
even go  as far  as to cause  an actual  expansion of the  dark matter
distribution. One  such process is dynamical  friction (e.g.  El-Zant,
Shlosman \& Hoffman  2001; Ma \& Boylan-Kolchin 2004;  Mo \& Mao 2004;
Tonini, Lapi \& Salucci 2006), which could be significant if disks are
built by  relatively big clumps.   Indeed, the physics of  gas cooling
suggests that disks may have formed out of clumpy, cold streams rather
than from  a smooth  cooling flow (Birnboim  \& Dekel 2003;  Maller \&
Bullock 2004; Dekel \& Birnboim 2006; Kaufmann \etal 2006; Keres \etal
2005).  We defer  a detailed study of the effects  of such cold infall
on the contraction of the halo  to a future paper.  Here we consider a
simple modification  of the BFFP adiabatic  contraction formalism that
allows  us,  with a  single  tunable  parameter,  to consider  reduced
contraction, no contraction, or even expansion.

Our method starts  from the BFFP formalism, according  to which a dark
matter particle initially (i.e., before the formation of the disk plus
bulge) at radius $\ri$ settles at a radius $\rf$, where
\begin{equation}
\label{blumac}
\rf \, \Mf(\rf) = \ri \, \Mi(\ri).
\end{equation}
Here   $\Mi(r)$  and  $\Mf(r)$   are  the   initial  and   final  mass
distributions.  If we assume that  initially the baryons have the same
(normalized) density distribution as the dark matter, then $\Mi(r)$ is
simply given  by the initial halo  profile (e.g. NFW).   For the final
mass distribution we have that
\begin{eqnarray}
\label{finalmassfunc}
M_{\rm f}(r_{\rm f}) & = & M_{\rm d}(r_{\rm f}) + M_{\rm b}(r_{\rm f})
+ M_{\rm  DM,f}(r_{\rm f}) \nonumber \\  & = & M_{\rm  d}(r_{\rm f}) +
M_{\rm b}(r_{\rm f}) + (1 - \mgal)M_{\rm i}(r_{\rm i})
\end{eqnarray}
Here  $\Md(r)$ is  the mass  of the  exponential disk  enclosed within
spherical  shells of radius  $r$, $\Mb(r)$  is the  similarly enclosed
mass of the Hernquist bulge,  and the second equality follows from the
assumption that  adiabatic contraction occurs  without shell crossing.
Equations~(\ref{blumac})   and~(\ref{finalmassfunc})  can   be  solved
iteratively  for  the  {\it  contraction factor}  $\Gamma(\ri)  \equiv
\rf/\ri$, which then allows for a computation of the mass distribution
of the contracted dark matter halo.

Our modification consists of simply defining the {\it actual} relation
between $\rf$ and $\ri$ as
\begin{equation}
\label{eqn:nu}
\rf = \Gamma^{\nu} \ri
\end{equation}
with  $\nu$ a  free  parameter: $\nu  =  1$ yields  the standard  BFFP
contraction, $\nu=0$ corresponds to no adiabatic contraction, and $\nu
<  0$ models  an expansion  of  the dark  matter halo.  As a  specific
example of an expansion model, $\nu=-1$ results in an expansion factor
that is equal to the contraction factor in the BFFP model.

Fig. \ref{fig:Vvir} shows the  impact of various adiabatic contraction
formalisms on  the ratio  of the total  circular velocity at  2.2 disk
scale lengths,  $\VII$, to the virial velocity,  $\Vvir$.  We consider
models  with an  effective  spin  parameter $\lamgal  =  0.048$ and  a
baryonic  mass  fraction  $\mgal  =  0.05$  (no  bulge  formation  is
considered  here).    We  apply  the   various  contraction  formalisms
described above  and compute the resulting $\VII/\Vvir$  as a function
of the  halo concentration parameter  $c$.  The three  vertical dashed
lines  (from  left to  right)  indicate  the  mean halo  concentration
expected for halos of mass  $\Mvir =10^{13}, 10^{12}, \, \& 10^{11} \,
h^{-1} M_{\odot}$ in the model of Bullock \etal (2001a).
 
Note that  with the standard  BFFP adiabatic contraction ($\nu  = 1$),
and  a  mean   halo  concentration  parameter  for  a   halo  of  mass
$\Mvir=10^{12}  h^{-1} M_{\odot}$, we  expect that  $\VII/\Vvir \simeq
1.7$,  while $\VII/\Vvir  \simeq 1.45$  without  adiabatic contraction
($\nu  = 0$).  In the  extreme case  where the  halo  is adiabatically
expanded ($\nu=-1$) the ratio is further lowered to $\VII/\Vvir \simeq
1.3$.   As an  illustration we  also plot  $V_{\rm  max}/\Vvir$, where
$V_{\rm max}$ is the maximum circular velocity of a NFW halo, which is
related to the halo concentration as
\begin{equation}
{V_{\rm max} \over \Vvir} \sim 
0.465 \sqrt{{c \over {\ln(1+c) - c/(1+c)}}}\,.
\end{equation}
Note that for the typical concentration of galaxy sized halos
$V_{2.2}\simeq V_{\rm max}$ if $\nu\simeq -1$.

The short  dashed (red) line in Fig.~\ref{fig:Vvir}  shows the results
for the adiabatic invariant $r \, M(\bar{r})$ proposed by Gnedin \etal
(2004).   As mentioned above,  this adiabatic  invariant results  in a
somewhat smaller overall contraction than the standard BFFP formalism.
Note that we can accurately model the Gnedin \etal formalism by simply
setting $\nu = 0.8$.  Not  only does this reproduce $V_{2.2}$ but also
the  {\it  shape} of  the  circular  velocity  profile.  Finally,  the
dot-short  dashed  (red) line  shows  the  results  for the  adiabatic
invariant $r \, V(\bar{r})$,  with $\bar{r}$ the orbit averaged radius
of Gnedin  \etal (2004).  This  adiabatic invariant accounts  for both
the eccentricity  of the orbits of  the dark matter  particles and for
the  (non-spherical) geometry of  the disk.   Note that  the resulting
$\VII/\Vvir$ is virtually indistinguishable from what one obtains with
the standard  BFFP formalism: taking account of  the non-sphericity of
the disk completely cancels the impact of non-circular orbits.

\subsection{Star Formation}
\label{sec:sf}

Disks  are made  up of  stars  and cold  gas. In  the highest  surface
brightness galaxies the gas fraction (defined as the ratio of cold gas
mass to total disk mass) is small, typically $\sim$10\%, so assuming a
pure stellar  disk is reasonable. However, the  gas fraction increases
with decreasing surface brightness  and luminosity (McGaugh \& de Blok
1997; Kannappan  2004) such that low surface  brightness galaxies have
comparable amounts of mass in cold  gas and stars. The data used here,
however, only  provides measurements of  the stellar disks,  while the
models describe  the distribution of  total baryonic matter  (cold gas
plus stars). To allow for  a proper comparison between models and data
we need a prescription for  computing the ratio between stars and cold
gas as function of radius in the galaxy.

Kennicutt (1989) has shown  that star formation is strongly suppressed
below a  critical surface  density, which can  be modeled by  a simple
Toomre stability criterion:
\begin{equation}
\Sigma_{\rm crit}(R) = {\sigma_{\rm gas}  \, \kappa(R) \over 3.36 \, Q
\, G}
\end{equation}
(Toomre  1964).   Here  $\kappa(R)=\sqrt{2}\frac{V(R)}{R}(1+\frac{d\ln
  V(R)}{d\ln    R})^{\frac{1}{2}}$   is   the    epicycle   frequency,
$\sigma_{\rm gas}$ is  the velocity dispersion of the  gas, and $Q$ is
the Toomre stability parameter.  Throughout we adopt $\sigma_{\rm gas}
= 6 \kms$ and $Q = 1.5$;  with this choice of parameters the radius at
which star formation is observed to truncate coincides with the radius
where $\Sigma_{\rm gas} = \Sigma_{\rm crit}$ (Kennicutt 1989).

We compute the total stellar mass as
\begin{equation}
\Ms = \Mb  + 2 \pi \int_0^{R_{ \rm  SF}} \left[\Sigma(R) - \Sigma_{\rm
    crit}(R)\right] \, R \, \dd R
\end{equation}
where $R_{\rm SF}$ is the star formation truncation radius, defined by
$\Sigma(R_{\rm SF})  = \Sigma_{\rm  crit}(R_{\rm SF})$.  Note  that we
thus assume that the bulge is made of stars entirely, and that all the
disk material  with a surface  density above the critical  density has
been   converted   into   stars.    Although  this   is   clearly   an
over-simplification  of the complicated  physics associated  with star
formation, there are several reasons  why this is probably reasonable. 
Firstly,  it is  a clear  improvement over  the assumption  that disks
consist of stars only, such as in the standard MMW model. Secondly, as
shown by van den Bosch (2000, 2001), this simple model (i) matches the
gas  mass fractions  of disk  galaxies  as function  of their  surface
brightness and luminosity, and (ii)  naturally leads to gas disks that
are more extended than stellar disks.
Finally,  more detailed,  `dynamic' models  for disk  formation, which
model  the actual  star  formation rate  using realistic,  empirically
motivated, prescriptions,  show that  the typical star  formation time
scale is short  compared to the time scale on  which the disk accretes
new gas.  Consequently, these models  indeed predict that the gas disk
has  been depleted  by  star  formation down  to  $\Sigma_{\rm gas}  =
\Sigma_{\rm crit}$  (van den Bosch 2001).   As we will  see later, the
inclusion  of a  star  formation threshold  density  proves a  crucial
ingredient in solving two problems of the standard MMW model.

\subsection{Conversion from Mass to Light}
\label{sec:ml}

In order to  compare our models with observations,  we need to convert
both the  stellar masses and the  stellar disk scale  lengths into the
observed $I$-band luminosities and scale lengths, respectively.

The  conversion from  mass to  light  is conventionally  done via  the
mass-to-light ratio, $\Upsilon \equiv  \Ms/L$.  Bell \& de Jong (2001)
showed that  $\Upsilon$ can be  estimated from optical  colors.  These
relations  have been  updated  by Bell  \etal  (2003a) and  Portinari,
Sommer-Larsen \& Tantalo (2004).   The main uncertainty in this method
is the normalization, reflecting  the unknown stellar IMF.  Additional
uncertainties  arise  from  details  related  to the  amount  of  dust
extinction  and the  star formation  histories.  Upper  limits  on the
normalization  can be  obtained  from maximal  disk  fits to  observed
rotation curves,  as shown in Bell  \& de Jong  (2001), although these
may still suffer from distance uncertainties.  More accurate estimates
from  rotation  curve  mass   modeling  are  hindered  by  well  known
degeneracies (e.g., van  Albada \etal 1985; van den  Bosch \etal 2000;
Dutton \etal 2005).

\begin{figure}
\begin{center}
\includegraphics[width=3.0in]{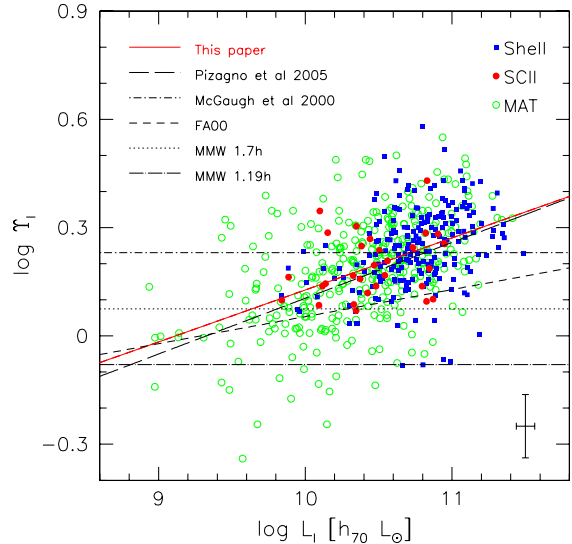}
\caption{I-band stellar mass-to-light ratio, $\YI$, versus
  I-band  luminosity. The  points  show $\YI$  estimated from  optical
  colors using the relations in Bell \etal (2003a).  These data form a
  sub-sample of the Courteau \etal (2006) data set: MAT (green), Shell
  (blue), and SCII  (red).  A typical 1$\sigma$ error  bar is given in
  the lower right corner. The solid red line gives the bi-weighted fit
  to these  data.  As a consistency  check the long  dashed line gives
  the  relation  computed using  the  mean  $(g-r)$ color  -luminosity
  relation  from  Pizagno \etal  (2005)  and the  $\Upsilon_i$-$(g-r)$
  color  relation  from Bell  \etal  (2003a).   We transform  $i$-band
  luminosities into  $I$-band luminosities  with $L_I$ =  $L_i$ -0.036
  (Courteau \etal 2006).  This relation is in excellent agreement with
  ours.   For comparison  we give  the $\YI=1.7$  estimate  of McGaugh
  \etal  (2000;  dot-dashed   line),  $\YI(L_I)$  relation  from  FA00
  (short-dashed line),  and the  adopted $\YI=1.7h$ (dotted  line) and
  the used $\YI=1.19h$ (dot-long dashed line) by MMW.}
\label{fig:ML-L}
\end{center}
\end{figure}

For the sub-sample  of our galaxies with optical  colors, we apply the
relations in  Bell \etal (2003a), to obtain  estimates of $\Upsilon_I$.
These are  based on a `diet'-Salpeter  IMF, introduced by  (Bell \& de
Jong  2001).   We have  verified  that,  assuming  the same  IMF,  the
relations  of Portinari \etal  (2004) give  very similar  results that
agree  to  within  $0.05$  dex.   In Fig.\ref{fig:ML-L}  we  plot  the
resulting $\YI$ as  a function of the $I$-band  luminosity.  The solid
red line shows  the bi-weighted orthogonal least squares  fit to these
points, and is given by
\begin{equation}
\label{eqn:MLL}
\log   \frac{\Upsilon_{I}}{[\Msun/\Lsun]}    =   0.172   +0.144   \log
\frac{\LI}{[10^{10.3}h_{70}^{-2}\Lsun ]} + \DeltaIMF ,
\end{equation}
The intrinsic scatter in this relation is estimated to be $\simeq 0.1$
dex (Bell \etal 2003a, Kauffmann  \etal 2003a).  We model this scatter
by  drawing  $\YI$  for  each  model disk  galaxy  from  a  log-normal
distribution  with  a mean  give  by  eq.~(\ref{eqn:MLL})  and with  a
scatter  $\sigma_{{\rm  ln}\Upsilon}$,  which   we  treat  as  a  free
parameter.  The other free parameter, $\DeltaIMF$, absorbs our lack of
knowledge about the IMF.  It  is equal to zero for the `diet'-Salpeter
IMF assumed here, while $\DeltaIMF \simeq +0.15$ for a Salpeter (1955)
IMF, and  $\DeltaIMF \simeq  -0.15$ for a  top-heavy IMF like  that of
Kennicutt (1983). See Portinari \etal (2004) for details.

As a consistency  check, we apply the Bell  \etal (2003a) relations to
the mean  $(g-r)$ color-luminosity  relation of Pizagno  \etal (2005).
The resulting $\Upsilon_{I}(\LI)$ is indicated as the long dashed line
in  Fig.~\ref{fig:ML-L} and agrees  with our  mean relation  to within
0.05 dex. For  comparison, we also show the  $\YI$ adopted by previous
studies.  The  short-dashed line shows  the relation adopted  by FA00;
the relatively  weak slope  of their $\Upsilon_{I}(\LI)$  relation was
derived  from a  comparison of  the  slopes of  $I$-band and  $H$-band
Tully-Fisher  relations, while  their  normalization is  based on  the
somewhat ad hoc  assumption that the disk contributes  $70$ percent to
the circular  velocity at $2.2$ disk scale  lengths.  Finally, McGaugh
\etal  (2000) and  MMW  both adopted  a  constant mass-to-light  ratio
(independent  of  luminosity).  The  value  adopted  by McGaugh  \etal
(2000), $\YI=1.7  (M/\LI)_{\odot}$, is  based on a  stellar population
synthesis  model with  a Salpeter  IMF, while  MMW based  their value,
$\YI=1.7  h  (M/\LI)_{\odot}$ on  the  sub-maximal  disk arguments  of
Bottema (1993).   Note, however,  that there is  an error in  MMW, and
that the actual value they  used is $\YI=1.19 h (M/\LI)_{\odot}$ (H.J.
Mo 2004, {\it private communication}).   Note that this value is small
compared to the  typical mass-to-light ratios shown here.   As we will
demonstrate  below,  this  has  important  implications  for  the  MMW
results.

\subsubsection{Disk Scale Lengths}
\label{sec:colgrad}

In order to compute the $I$-band scale lengths of our model disks, for
direct  comparison with  the data  discussed in  \S\ref{sec:data}, we
proceed as follows.

For each galaxy  we first compute the stellar  surface density profile
of the disk, i.e., $\Sigma_{*}= \Sigma(R) - \Sigma_{\rm crit}(R)$ with
$0 \leq R \leq R_{\rm SF}$, which we fit with an exponential disk over
the radial range $0.15 R_{\rm SF} \leq R \leq 0.80 R_{\rm SF}$.

In theoretical models of disk galaxy formation, disks form inside out.
This  results  in color  gradients,  with  progressively larger  scale
lengths when  going from  stellar mass to  $K$-band light  (assumed to
closely  trace  the  underlying  stellar  mass)  to  $B$-band  light.  
Observations show a  similar trend with $R_{R}/R_{H}=1.17$ (MacArthur,
Courteau, \&  Holtzman 2003),  and $R_{I}/R_{K}=1.13$ (de  Jong 1996),
where subscripts  denote the photometric  band.  To account  for these
color gradients we convert our  stellar disk scale lengths to $I$-band
scale lengths  using $\log R_{I}  = \log \Rs  + 0.05$, with  $\Rs$ the
scale length  of the exponential fit  to the model  stellar disk. Note
that this assumes the scale length in stellar mass, $\Rs$, is equal to
the scale  length in  $K$-band light, $R_K$,  and thus is  probably an
underestimate of the true correction.

\subsection{Sampling Strategy}
\label{sec:sampling}

In order to pursue a meaningful  comparison of our model $VL$ and $RL$
relations  with  the  observations,  we construct  samples  of  random
realizations  of model galaxies.   Each model  galaxy is  specified by
four  parameters; the  virial  mass  of the  halo,  $\Mvir$, the  halo
concentration  parameter, $c$,  the  effective spin  parameter of  the
galaxy,  $\lamgal$, and  the  baryonic mass  fraction  of the  galaxy,
$\mgal$.

Previous studies  (e.g., MMW; Somerville  \& Primack 1999;  Cole \etal
2000; van  den Bosch 2000;  Firmani \& Avila-Reese 2000;  Croton \etal
2006) have  assumed  that $\lamgal  =  \lambda$.   However, there  are
numerous reasons why  the spin parameter of the  disk may be different
from that of its dark  matter halo (see \S\ref{sec:lowspin}).  In this
paper we assume that $\lamgal$ follows a log-normal distribution:
\begin{equation}
p(\lamgal)\,\dd\lamgal =  \frac{1}{\sigma_{\ln\lambda}\sqrt{2 \pi}} 
\exp \left[-\frac{\ln^2(\lamgal/\lambar_{\rm gal})}
{2\sigma_{\ln\lambda}^2} \right] \frac{\dd\lamgal}{\lamgal},
\end{equation}
and  we treat  $\lambar_{\rm gal}$  and $\sigma_{\ln\lambda}$  as free
parameters. If  $\lamgal \simeq \lambda$ we  expect that $\lambar_{\rm
  gal} \simeq  0.042$ and $\sigma_{\ln\lambda} \simeq  0.5$, which are
the values obtained for dark matter haloes from cosmological numerical
simulations (Bullock  \etal 2001b). In  what follows we  will consider
these as our fiducial values.

The  concentration parameter,  $c$,  is strongly  correlated with  the
halos mass  accretion history (Wechsler  \etal 2002; Zhao  \etal 2003;
Li, Mo \& van den Bosch 2005),  and thus depends on both halo mass and
cosmology.   Bullock  \etal (2001a)  and  Eke,  Navarro, \&  Steinmetz
(2001) present   analytical  models,   calibrated   against  numerical
simulations, for the computation of  a mean concentration given a halo
mass and cosmology.  In what follows we use the model by Bullock \etal
(2001a).   At a  given halo  mass,  the halo  concentrations follow  a
log-normal distribution with a scatter $\sigma_{\ln c}=0.32$ (Wechsler
\etal 2002).  However, as for the spin parameter, the mean and scatter
are different for  the subset of halos without a  recent major merger. 
As shown by  Wechsler \etal (2002), the scatter  becomes smaller while
the  mean increases.   To be  able to  account for  this,  we consider
$\sigma_{\ln  c}$ a free  parameter and  compute $c(\Mvir)$  using the
relation by Bullock  \etal (2001a) but multiplied by  a free parameter
$\eta_{\rm c}$.

Unlike  for $c$  and $\lambda$,  very  little is  known regarding  the
baryonic  mass fractions, $\mgal$,  of galaxies.  We account  for this
limitation by modeling the mean as
\begin{equation}
\label{eqn:mMvir}
\overline{m}_{\rm gal}(\Mvir) = \mgalo  \left( {\Mvir \over 10^{11.5} h^{-1}
\Msun} \right)^{\alpha_{\rm m}}
\end{equation}
with $\mgalo$ and $\alpha_{\rm m}$  two free parameters.  The value of
$\alpha_{\rm m}$  is related to  the relative efficiencies  of cooling
and feedback processes. Typically,  cooling results in $\alpha_{\rm m}
< 0$  while feedback results in  $\alpha_{\rm m} > 0$  (e.g., Dekel \&
Silk 1986; van den Bosch 2002).   Finally, to allow for scatter in the
above relation, we assume that,  at fixed halo mass, $\mgal$ follows a
log-normal distribution  with scatter $\sigma_{\ln m}$,  which we also
consider a free parameter.

To  construct $VL$ and  $RL$ relations  for our  models we  proceed as
follows. We uniformly sample a  range in $\log{\Mvir}$. For each halo,
we  then  draw  values  for  $c$, $\lamgal$,  and  $\mgal$  using  the
log-normal  distributions described  above.  We  then iterate  until a
solution  for  the  disk-to-bulge  ratio is  found,  taking  adiabatic
contraction into account, and compute the resulting galaxy luminosity,
$L_I$, disk scale length $R_I$, (both in the $I$-band), as well as the
circular velocity,  $V_{2.2}$, of the  model galaxy at $2.2  R_I$.  To
mimic  the observational  errors we  add to  each of  the  three model
observables  $\log  V_{2.2}$,  $\log\LI$,  and  $\log\RI$  a  Gaussian
deviate  with a  dispersion  that reflects  the typical  observational
errors.   The uniform  sampling  of halos  in  $\log\Mvir$ results  in
approximately  uniform  sampling in  $\log\LI$,  while the  luminosity
sampling of the data is approximately log-normal with a low luminosity
tail.   To allow  for  a fair  comparison  with the  data  we need  to
reproduce  the  observed  luminosity  sampling.   We do  this  with  a
Monte-Carlo  technique which  accepts  model galaxies  such that  they
reproduce  the   observed  luminosity  distribution.    Finally,  when
including bulge formation, we tune $\beta_{\rm crit}$ so as to roughly
reproduce  the observed  distribution of  bulge-to-disk ratios  of our
data sample.

\subsection{Overview of Model Parameters}
\label{sec:param}

A  list of the  free model  parameters with  their fiducial  values is
given   in  Table~1.   The   fiducial  value   $\beta_{\rm  crit}=1.0$
corresponds to a model without  bulge formation.  

\begin{deluxetable}{lrl}
\tabletypesize{\small}  
\tablecaption{Overview of Model Parameters and their Fiducial Values}
\tablewidth{0pt} 
\tablehead{  
\colhead{Parameter} & \colhead{symbol} & \colhead{fiducial value} }  
\startdata 
Median effective spin parameter & $\overline{\lambda}_{\rm gal}$ & 0.042\\   
Concentration parameter normalization & $\eta_{\rm c}$       & 1.0  \\ 
Adiabatic contraction parameter      & $\nu$                & 1.0  \\ 
Toomre stability parameter           & $Q$                  & 1.5  \\ 
Bulge  formation threshold           & $\beta_{\rm crit}$   & 1.0  \\ 
Bulge formation exchange parameter   & $\fx$                & 0.5  \\ 
Galaxy mass fraction normalization   & $\mgalo$             & 0.05 \\ 
Galaxy mass fraction slope           & $\alpham$            & 0.0  \\  
Mass-to-light ratio normalization    & $\DeltaIMF$          & 0.0  
\enddata
\end{deluxetable}

In addition to these
model parameters, there are  four parameters that describe the amounts
of scatter in the  log-normal distributions of $c$, $\lamgal$, $\mgal$
and $\YI$, and which we also treat as free parameters in what follows.
   

\section{Comparison between Models and Data}
\label{sec:convert}

In  order to  gain some  insight  into the  origin of  the slopes  and
zero-points of the $VL$ and  $RL$ relations, we start by considering a
set  of   simplified  models  without   bulge  formation  ($\beta_{\rm
  crit}=1.0$), and without scatter ($\sigma_{{\rm ln}c} = \sigma_{{\rm
    ln}\lambda} = \sigma_{{\rm  ln}m} = \sigma_{{\rm ln}\Upsilon}=0$). 
More complete models, including  both bulge formation and scatter will
be presented in \S\ref{sec:modelscatter}.

\subsection{Median Parameter Relations}
\label{sec:medparrel}

\subsubsection{Slopes}
\label{sec:slopes}

\begin{figure*}
\begin{center}
\includegraphics[width=6.0in]{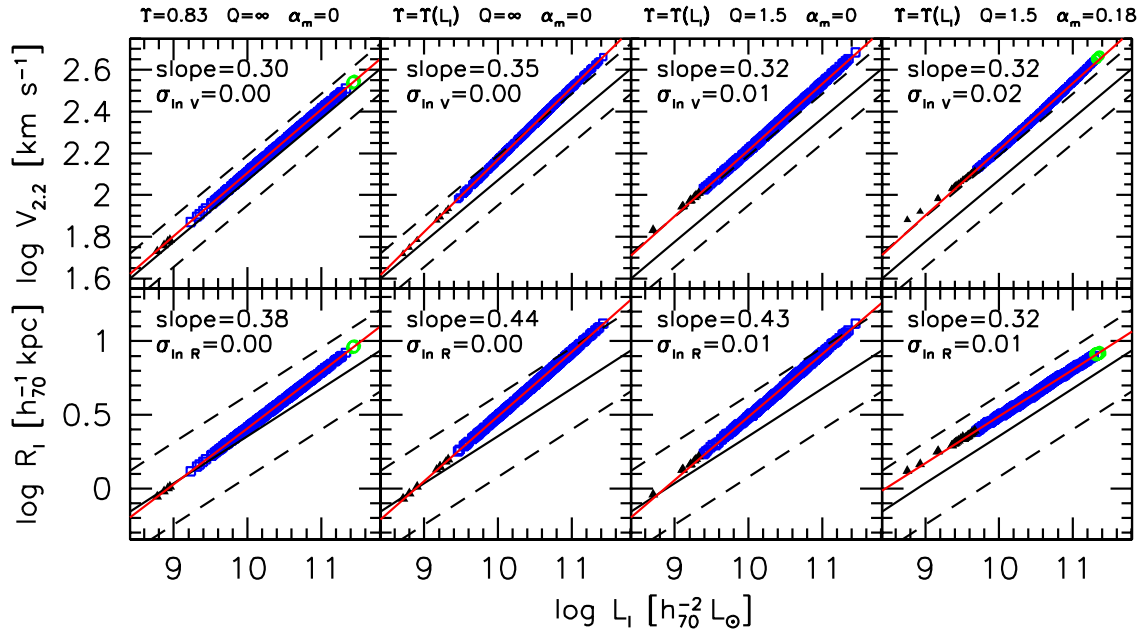}
\caption{$VL$ and $RL$ relations using our fiducial mean $c$, $\lamgal$,
  and  $\mgal$  parameters and  different  values  of  $\YI$, $Q$  and
  $\alpham$.   The black  solid and  dashed  lines show  the mean  and
  $2\sigma$ scatter of the observations. The  red line is a fit to the
  model galaxies, whose slope is given  in the top left corner of each
  box.   The  color  of  the   points  corresponds  to  $\muI$  as  in
  Fig.~\ref{fig:VLR_ALL}.  The model on the  far left is the MMW model
  with  $\Upsilon_I=0.83$,  $Q=\infty$   (a  pure  stellar  disk)  and
  $\alpham=0$. The  model in  the left middle  only differs  from this
  model in the $\YI$, which we assume follows Eq.~\ref{eqn:MLL}.  This
  steepens  the slopes  of the  $VL$  and $RL$  relations making  them
  incompatible  with the data.   The right  middle model  includes the
  star  formation threshold,  which  results in  a  gas fraction  that
  increases  with decreasing  luminosity, and  hence a  shallower $VL$
  slope, in agreement  with the data. The effect on  the $RL$ slope is
  not as strong because as  the luminosities are decreased, so are the
  stellar  disk  scale  lengths.    Notice  that  the  star  formation
  threshold has introduced a small curvature to both the $VL$ and $RL$
  relations.  In  order to  match the $RL$  slope we  set $\alpha_{\rm
    m}=0.18$   (far  right  panel).    Although  $\mgal$   now  varies
  significantly with luminosity, the  only effect on the $VL$ relation
  is to increase the curvature  slightly. While this model matches the
  slope of the $VL$ and $RL$  relations it fails to reproduce the zero
  points.}
\label{fig:VLR8-M0}
\end{center}
\end{figure*}

We  start by  reproducing the  models of  MMW. To  this extent  we set
$\alpha_{\rm m}=0$,  $Q=\infty$ (i.e., entire disk is  made of stars),
and  $\YI=0.83$ (i.e.   $1.19 \times  0.7$).  As  can be  seen  in the
left-hand panels  of Fig.~\ref{fig:VLR8-M0} this  model reproduces the
slope of the $VL$ relation,  but predicts a somewhat steeper slope for
the $RL$ relation than observed.  The deviations of the slopes of both
the $VL$ and $RL$ relations from  the virial value of $1/3$ are due to
the  non-homology  of the  dark  matter  halos  (i.e., the  $c(\Mvir)$
relation).  Our deviations from the pure virial equations are somewhat
larger  than in  MMW.  This  owes to  the fact  that we  use  the halo
concentration model  of Bullock \etal  (2001a), which predicts  a much
stronger mass dependence  than the NFW model used  by MMW. The Bullock
\etal (2001a) model also predicts  higher $c$ than the NFW model. This
accounts for the slight discrepancy between the $VL$ zero point of our
model and the data.

Panels in  the second column  from the left  show the same  model, but
with the more realistic $\Upsilon(L)$ relation of eq.~(\ref{eqn:MLL}).
Note  that the  slopes of  both the  $VL$ and  $RL$ relations  are now
significantly  steeper  than  the   data.   In  the  third  column  of
Fig.~\ref{fig:VLR8-M0},  we include  the star  formation  threshold to
separate the disk into stars and gas.  Since lower luminosity galaxies
have higher  gas-to-stellar mass ratios this flattens  the $VL$ slope,
bringing it back in agreement with the data\footnote{We emphasize that
  the luminosity  dependence of the  gas-to-stellar mass ratio  is the
  key ingredient to  this success, not the mechanism  by which this is
  achieved.  Therefore  any argument against $\Sigma_{\rm  crit}$ as a
  physical  threshold for star  formation is  not relevant  here.  All
  that matters is  that our separation of disk mass  in stars and cold
  gas reproduces the gas-to-stellar mass  ratios as a function of disk
  luminosity and surface brightness, as  is the case for our models.}. 
The  star formation  threshold also  results in  smaller  stellar disk
scale lengths  compared to the baryonic disk  scale lengths.  However,
this  reduction is  approximately  canceled out  by  the reduction  in
luminosity  so   that  the  $RL$   slope  and  zero  points   are  not
significantly affected  by the star formation  threshold.  Finally, in
order to match the $RL$ slope  we tune $\alpham$ to match the slope of
the $RL$  relation.  This requires $\alpham=0.18$, so  that the galaxy
mass  fraction,  $\mgal$,  increases  with increasing  halo  mass,  as
expected,  for example,  for simple  supernova feedback  models (e.g.,
Dekel \& Silk 1986; van den  Bosch 2002; Dekel \& Woo 2003).  As shown
in the right-hand panels  of Fig.~\ref{fig:VLR8-M0} this model roughly
matches  the slopes  of both  the $VL$  and $RL$  relations,  and will
hereafter serve as our reference model.

Although the  introduction of a  non-zero $\alpham$ causes  the galaxy
mass fractions, $\mgal$,  to systematically vary by a  factor $\sim 3$
over the  luminosity range probed, this does  not significantly impact
the slope of the $VL$ relation. This owes to the fact that variance in
$\mgal$ scatters galaxies mainly  along the $VL$ relation: an increase
in $\mgal$ makes the galaxy more luminous but simultaneously increases
its  rotation velocity  (see also  Navarro \&  Steinmetz  2000).  This
important   aspect,    which   we   discuss   in    more   detail   in
\S\ref{sec:modelscatter}  below,  implies  that  we  can  simply  tune
$\alpham$ and  $\mgalo$ to fit  the slope and  zero point of  the $RL$
relation, without (strongly) affecting the $VL$ relation.

\subsubsection{Zero points}
\label{sec:zeropoint}

Having explored how some of  our model parameters impact the slopes of
the $VL$ and $RL$ relations, we  now turn to the zero points, which we
define as  the values  of $V$ and  $R$ at  the mean luminosity  of our
data;  $\log\LI=10.3$.  For  our reference  model  with $\alpham=0.18$
which matches  the slopes  of the $VL$  and $RL$ relations,  the model
galaxies are both too large (by $\sim 35$ percent) and rotate too fast
(by $\sim  40$ percent).   This corresponds to  a $2\sigma$  offset in
terms of the  observed scatter in the $VL$ relation,  and thus an even
more significant offset in terms of the actual uncertainty in the $VL$
zero point.

The inability of  CDM based galaxy formation models  to match the $VL$
zero point is a generic  problem that has been identified in numerical
simulations  (Elizondo \etal  1999;  Navarro \&  Steinmetz 2000,  Eke,
Navarro \& Steinmetz 2001; Sommer-Larsen \etal 2003), in semi-analytic
models  (e.g.,  Mo  \&  Mao  2000;  van den  Bosch  2000,  Firmani  \&
Avila-Reese  2000; Cole  \etal 2000;  Benson \etal  2003) and  in halo
occupation models (Yang, Mo, \& van den Bosch 2003).  In particular no
model has  been able to  simultaneously match the  luminosity function
and  the  zero  point  of  the $VL$  relation,  using  standard  \LCDM
parameters.   Cases that  claim an  agreement either  assume  that the
observed rotation velocity, $V_{\rm  obs}$, is equal to $\Vvir$ (e.g.,
Somerville \& Primack  1999), or to $V_{\rm max}$  (e.g.  Croton \etal
2006).  In both cases, the effect of the baryons on the rotation curve
is  completely   ignored.   As  shown   in  Fig.~\ref{fig:Vvir},  when
adiabatic  contraction of  the dark  matter  halo and  the disk's  own
contribution to $V_{\rm obs}$ are properly accounted for, one predicts
that  $V_{\rm obs}/\Vvir\simeq  1.7$ for  a  typical halo  in a  \LCDM
cosmology.  Models  that fail to  take these effects into  account can
therefore not be  used for a meaningful comparison  with the data (see
also Navarro  \& Steinmetz  2000; van den  Bosch 2000).   For example,
Croton  \etal  (2006) model  $V_{\rm  obs}$  as  the maximum  circular
velocity of  a NFW  halo, $V_{\rm max}$.   As shown by  the dot-dashed
curve (labelled $V_{\rm max}/\Vvir$) in Fig.~\ref{fig:Vvir}, this more
or less corresponds to halo expansion with $\nu=-1.0$.  We will return
to the implications of this in \S\ref{sec:lowspin}.

In principle, there are a number  of different ways in which one might
envision solving the $VL$ zero point problem:
\begin{enumerate}
  
\item Lower stellar mass-to-light ratios. This trivially increases the
  luminosity of the  model galaxies at a fixed  rotation velocity. For
  our reference model we need  to lower $\Upsilon$ by $\simeq0.5$ dex. 
  However, the  most that one can  justify based on  realistic IMFs is
  $\Delta_{\rm  IMF} \simeq  -0.2$ dex  (corresponding to  a Kennicutt
  IMF), more than a factor of 2 smaller than what is required.
  
\item Lower halo concentrations.  This results in a lower $V_{2.2}$ at
  a given  $L$. Halo  concentrations can be  lowered by,  for example,
  decreasing the cosmological  parameters $\OmegaM$ and/or $\sigma_8$. 
  A reduction  of the power spectrum  on small scales  also results in
  lower halo  concentrations (Zentner \& Bullock  2002).  However, for
  our reference  model to  match the $VL$  zero point we  need $\eta_c
  \simeq 0.3$ which is difficult to reconcile with current constraints
  on cosmological parameters and the matter power spectrum.
  
\item Modify adiabatic contraction.   As discussed in \S\ref{sec:ac} a
  more realistic treatment of  adiabatic contraction than the standard
  BFFP  formalism  is  unlikely  to  have any  significant  impact  on
  $V_{2.2}$.   However, if  we simply  turn off  adiabatic contraction
  (i.e.   $\nu=0$), $V_{2.2}$ will  be lowered  by $\simeq  20\%$.  An
  even   stronger  reduction  can   be  accomplished   by  considering
  expansion, modelled by setting $\nu < 0$ in Eq.~(\ref{eqn:nu}).

\end{enumerate}

\begin{deluxetable}{cccccccccccc}
\tabletypesize{\small}         
\tablecaption{Model Parameters \label{tbl:modpar}}
\tablewidth{0pt} 
\tablehead{\colhead{Model} & \colhead{$\Delta_{IMF}$} & \colhead{$\nu$} & 
\colhead{$\eta_c$} & \colhead{$Q$} & 
\colhead{$\overline{\lambda}_{\rm gal}$} & 
\colhead{$\mgalo$} & \colhead{$\alpha_{m}$} & \colhead{$\beta_{\rm crit}$} &  
$\fx$  }
\startdata 
  I & -0.40 &$\,$ 0.8 & 0.80 & 1.5 & 0.042 & 0.06 & 0.25 & 0.66 & 0.25\\
 II & -0.20 &$\,$ 0.8 & 0.50 & 1.5 & 0.042 & 0.10 & 0.25 & 0.75 & 0.25\\  
III & -0.20 &$\,$ 0.0 & 0.80 & 1.5 & 0.042 & 0.09 & 0.25 & 0.80 & 0.25\\
 IV & -0.20 &-1.0 & 1.00 & 1.5 & 0.023 & 0.03 & 0.30 & 0.85 & $\;$0.25
\enddata
\end{deluxetable}

Based on these possibilities we  construct three models that match the
$VL$ and $RL$ zero points  and slopes.  The parameters of these models
are listed in  Table~2. In each of these models  we tune the parameter
$\mgalo$ and  $\alpham$ to  fit the $RL$  zero point and  slope.  Note
that these  models should  only be considered  specific examples  of a
more extended  parameter space,  which we describe  in more  detail in
\S\ref{sec:lowspin}.   These  models  mainly  serve to  highlight  the
various possible  solutions to the  $VL$ zero point  problem discussed
above.

In model~I we consider elements from all three modifications discussed
above.   We reduce  the  halo  concentrations by  20\%  (i.e.  we  set
$\eta_c=0.8$), which corresponds to changing $(\OmegaM,\sigma_8)$ from
$(0.3,0.9)$ to $(0.25,0.8)$, as advocated by van den Bosch, Mo \& Yang
(2003) and  which is  in agreement  with the  third year  WMAP results
(Spergel \etal 2006).  In addition,  we use the adiabatic invariant of
Gnedin \etal  (2004) which  corresponds to $\nu  = 0.8$.   Finally, we
adjust   the  stellar   mass-to-light  ratios   until  we   match  the
zero-points, which  requires $\Delta_{\rm IMF} =  -0.4$.  As discussed
above,  such a  large reduction  in $\YI$  implies  an unrealistically
top-heavy  IMF.   This model  should  therefore  be  considered as  an
illustration  only.  Because  of the  lower $c$  and the  reduced halo
contraction,  the  resulting disk  galaxies  are  larger  than in  our
reference  model.   We  counter-balance  this  by  slightly  increasing
$\mgalo$ from our fiducial  value of $0.05$ to $0.06$.  Alternatively,
we could  have matched the  increase in disk scale  lengths by
decreasing $\overline{\lambda}_{\rm  gal}$: such models  are discussed
in \S\ref{sec:lowspin}.

In model~II  we restrict  $\Delta_{\rm IMF}$ to  $-0.2$, which  is the
most we  can accommodate  with realistic IMFs,  and we match  the $VL$
zero point  by lowering $c$ by  50\%.  As shown by  Zentner \& Bullock
(2002),  such  a  large   reduction  in  halo  concentrations  can  be
reconciled with a power spectrum with a running spectral index that is
still consistent with the WMAP data (Spergel \etal 2003).

Finally,  in  model~III  we  simply  turn  off  adiabatic  contraction
($\nu=0$). This model  is able to match the $VL$  and $RL$ zero points
with  the  standard  $c(M_{\rm  vir})$  for  a  \LCDM  cosmology  with
$\Omega_m = 0.25$ and $\sigma_8  = 0.8$ (i.e., $\eta_{\rm c}=0.8$) and
with a realistic IMF ($\Delta_{\rm IMF}=-0.2$).

\subsection{Models with Scatter}
\label{sec:modelscatter}

Having identified models that  can simultaneously match the slopes and
zero  points of  the $VL$  and  $RL$ relations,  we now  focus on  the
scatter in both relations.  The  observed scatter in the $VL$ and $RL$
relations at  fixed $\LI$  are $\sigma_{\ln V}=0.13$  and $\sigma_{\ln
  R}=0.32$   respectively.    The   observational  uncertainties   are
estimated   to   be   $\sigma_{\ln  V}\simeq0.08$,   $\sigma_{\rm   ln
  L}\simeq0.10$,      and     $\sigma_{\ln      R}\simeq0.14$     (see
\S\ref{sec:datacorr}).  Subtracting these in quadrature (with an error
weighted scheme) from the observed scatter leaves an intrinsic scatter
of  $\sigma_{\ln V}=0.12$  and $\sigma_{\ln  R}=0.28$.  Note  that our
error  weighted subtraction  results  in a  slightly larger  intrinsic
scatter than obtained from a straight quadratic subtraction.

The scatter  in our models  originates from four sources:  $\YI$, $c$,
$\lamgal$, and $\mgal$.  Table~3  lists the expected amount of scatter
in each of these  four variables, and Fig.\ref{fig:VLR-mcl8} shows the
affect of  these sources of  scatter on the  $VL$ and $RL$  relations. 
Lower $c$,  higher $\lamgal$, and  lower $\mgal$ all result  in larger
scale  lengths, and  lower  stellar masses  and luminosities.   Higher
$\YI$ result in  lower luminosities, and higher effective  (i.e.  at a
fixed luminosity)  scale lengths.   The fact that  scatter in  $c$ and
$\lamgal$  also effects  the luminosity  is entirely  due to  the star
formation  threshold  and  results  in a  significantly  reduced  $VL$
scatter due to $\lamgal$.  As already discussed in \S\ref{sec:slopes},
scatter in  $\mgal$ moves galaxies approximately parallel  to the $VL$
relation,  but  only  for  intermediate  values  of  the  galaxy  mass
fraction:  for large  $\mgal$ scatter  results in  $V\propto L^{0.5}$,
while for  sufficiently low values of $\mgal$  the corresponding slope
approaches zero.

Fig.~\ref{fig:VLR-mcl8} also  yields useful  insight on the  effect of
the  various  parameters on  the  zero points  of  the  $VL$ and  $RL$
relations. The most effective way to  change the $VL$ zero point is to
change  the average halo  concentration, $c$,  or the  average stellar
mass-to-light ratio,  $\YI$. The zero  point of the $RL$  relation, on
the other hand,  is most easily changed by a  modification of the mean
spin  parameter,  $\lamgal$,  or  the average  galaxy  mass  fraction,
$\mgal$.

\begin{deluxetable}{lcccc}
  \tablecaption{Sources  of   intrinsic  scatter  \label{tbl:scatter}}
  \tabletypesize{\small}   \tablewidth{0pt}   \tablehead{\colhead{}  &
    \colhead{$\sigma_{\ln\,\Upsilon_I}$} & \colhead{$\sigma_{\ln\,c}$}
    &                \colhead{$\sigma_{\ln\,\lambda}$}               &
    \colhead{$\sigma_{\ln\,\md}$}} \startdata
  Maximum  & 0.32 & 0.40 & 0.25 & 0.55$\;\;\,$ \\
  Predicted  & 0.23 & 0.32 & 0.50 &   -  \\
  Adopted    &   0.23    &    0.23   &    0.25    &   0.00    \enddata
  \tabletypesize{\scriptsize}  \tablecomments{The first row  gives the
    maximum  scatter in  $\YI$, $c$,  $\lamgal$, and  $\mgal$  that is
    consistent with the  $VL$ and $RL$ scatter.  The  second row gives
    the  predicted scatter,  for $c$  and $\lamgal$  from cosmological
    simulations, for $\YI$ from observations.  The third row gives the
    scatter adopted  in models  I-III in order  to match the  $VL$ and
    $RL$ scatter simultaneously while including bulge formation.}

\end{deluxetable}

\subsubsection{Constraining the amount of scatter}
\label{sec:limscatter}

\begin{figure*}
\begin{center}
\includegraphics[width=6.0in]{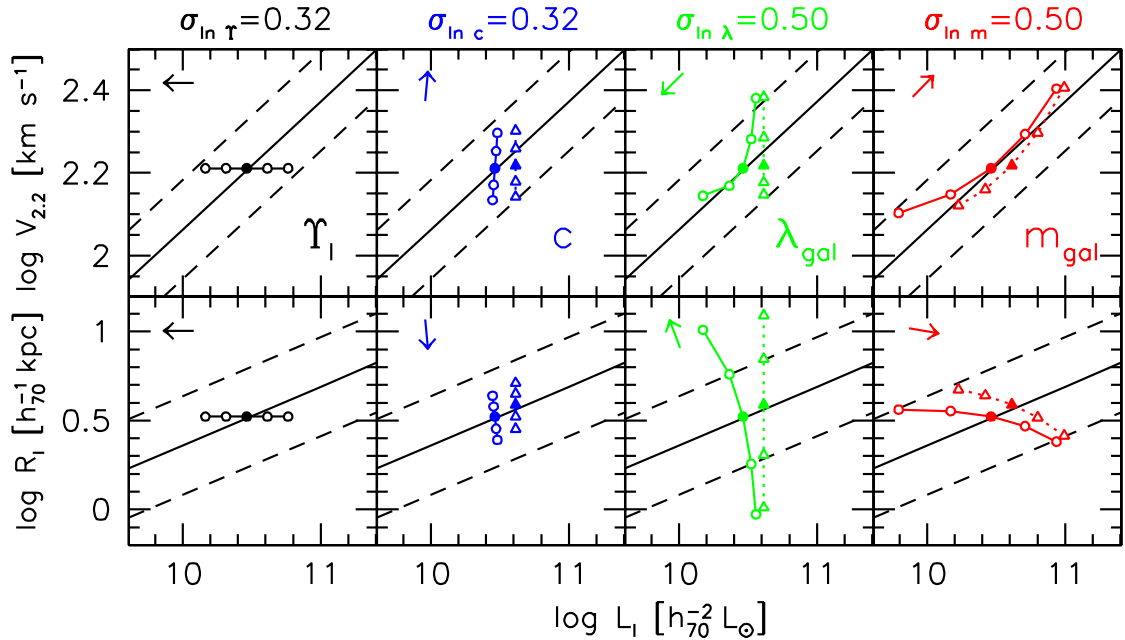}
\caption{Contribution of scatter in $\YI$, $c$, $\lamgal$, and
  $\mgal$  to the  I-band  $VL$  and $RL$  scaling  relations for  our
  reference  model with $\Mvir  = 3\times  10^{11}\,h^{-1}\,\Msun$ and
  $\DeltaIMF=-0.5$  (to match the  zero points).   Models with/without
  the  star formation  threshold  are given  by circles/triangles.  The
  solid symbol shows  the mean parameter model, the  open symbols show
  the models with  $\pm 1$ \& $2\sigma$ scatter  in the parameter. The
  arrow in the top left of each panel shows the direction in which the
  parameter increases.   The observed  mean and $2\sigma$  scatter are
  given by solid and dashed lines, respectively.  See text for further
  details.}
\label{fig:VLR-mcl8}
\end{center}
\end{figure*}

To determine the maximum amount  of scatter allowed for each parameter
we run  models with log-normal scatter  in one of  the four parameters
only.  We increase the scatter until we reach the intrinsic scatter of
either the $VL$ or $RL$ relation.  The resulting limits on the amounts
of scatter  in $c$,  $\lamgal$, $\mgal$, $\YI$,  are very  similar for
each  of the three  models discussed  above.  In  Table~3 we  list the
representative    values   thus    obtained.     As   expected    from
Fig.\ref{fig:VLR-mcl8},  the limits for  $\YI$ and  $c$ come  from the
$VL$ relation, while  those for $\lamgal$ and $\mgal$  owe to the $RL$
relation. Note that the {\it expected} amounts of scatter (also listed
in Table~3)  in $\YI$  and $c$ are  smaller, and  therefore consistent
with,  their respective  maximum amounts.   In  the case  of the  spin
parameter, however,  the expected  amount of scatter  is a  factor two
larger than the maximum amount allowed by the intrinsic scatter in the
$RL$ relation.  This  can have a number of  important implications, of
which we consider the following three:
\begin{enumerate}
  
\item Due to surface brightness  selection effects our data is missing
  the lowest and/or highest surface brightness galaxies. Although this
  is   certainly   possible  to   some   extent,   gauging  from   the
  incompleteness corrections  of de Jong  \& Lacey (2000)  we conclude
  that these effects are likely to be small (Courteau \etal 2006).
  
\item  Disk  galaxies  acquire  a narrower  distribution  of  specific
  angular  momentum  than  their  host  halos (i.e.,  the  scatter  in
  $\ln\lamgal$ is smaller than that in $\ln\lambda$).  This can occur,
  for  example,   through  the  redistribution   of  angular  momentum
  associated with  bulge formation.  We discuss such  a possibility in
  the next section.
 
\item  Disk  galaxies  form in  a  sub-set  of  halos with  a  smaller
  $\sigma_{\ln  \lambda}$.  Interestingly,  this is  expected  if disk
  galaxies form mainly in those halos that have not experienced recent
  major mergers (D'Onghia \& Burkert  2004).  In addition to a smaller
  scatter in both  $\lambda$ and $c$, this subset of  halos also has a
  significantly  smaller mean  spin parameter,  and possibly  a higher
  mean  $c$  (Wechsler  \etal  2002).   We discuss  such  a  model  in
  \S\ref{sec:lowspin}.
  
\end{enumerate}

Having determined the maximum amounts  of scatter allowed by the data,
we now proceed to set their actual values. Since scatter in $\md$ does
not contribute much  scatter to the $VL$ relation,  and the scatter in
the  $RL$  relation  is  already  over-budgeted because  of  the  spin
parameter,  we set  $\sigma_{{\rm  ln}m}  = 0$  in  what follows.   In
addition, we  set the scatter  in the stellar mass-to-light  ratios to
the  expected amount, i.e.,  $\sigma_{{\rm ln}\Upsilon}=0.23$.   If we
also set $\sigma_{{\rm  ln}c}$ to its predicted value  of $0.32$, then
the scatter in the $VL$ relation is already larger than observed, even
without any  scatter in  $\lamgal$.  We therefore  adopt $\sigma_{{\rm
    ln}c}=0.23$, which is the value  predicted for halos that have not
experienced any recent major  mergers (Wechsler \etal 2002).  Finally,
we tune the  scatter in $\lamgal$ until we  match the observed scatter
in  the $RL$  relation, which  yields  $\sigma_{{\rm ln}\lambda}=0.25$
(when including bulge formation,  see below). Throughout this paper we
treat $c$, $\YI$, $\mgal$  and $\lamgal$ as independent variables.  In
reality, the scatter in some of these parameters may be correlated. In
Appendix~\ref{sec:modcor}  we discuss  possible correlations,  and how
they impact on our results.

\begin{figure*}
\begin{center}
\includegraphics[width=6.0in]{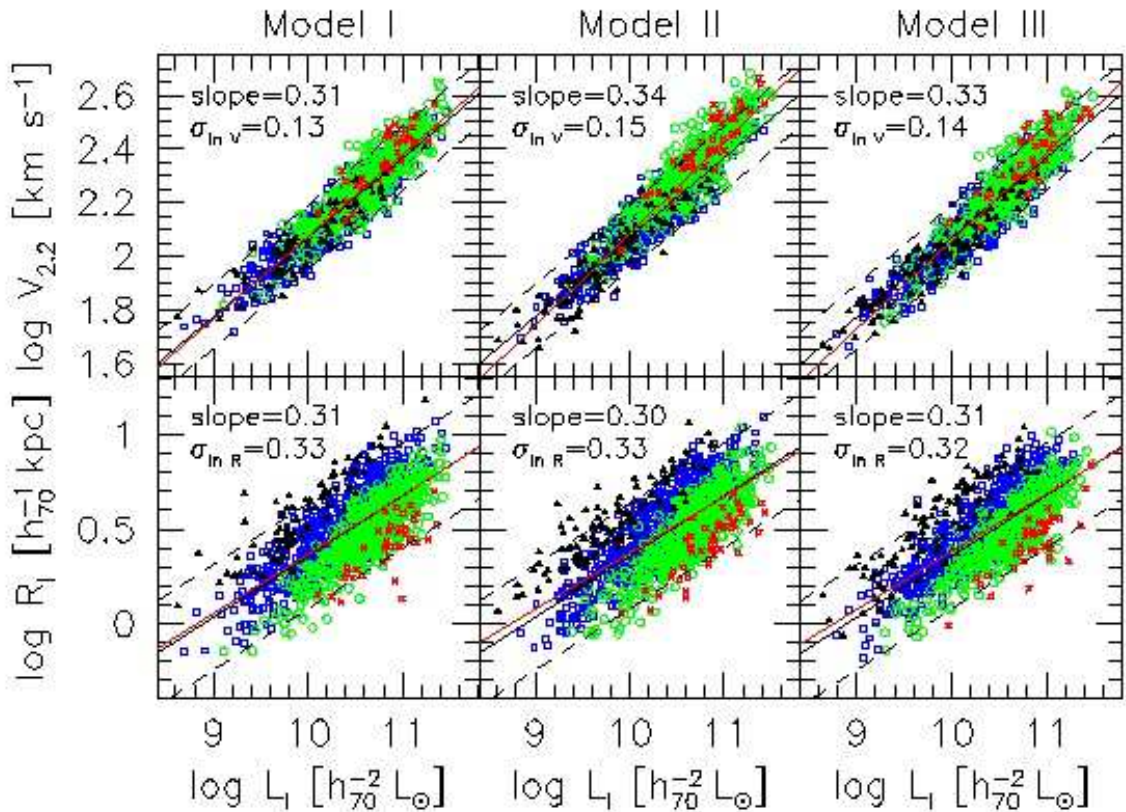}
\caption{I-band $VL$ and $RL$ scaling relations for our three models with
  bulge   formation  and   scatter  in   $c$,  $\lamgal$,   $\YI$  and
  observational errors.   The parameters of these models  are given in
  Table~2.  The  black solid and dashed  lines show the
  mean and $2\sigma$ scatter of the data, respectively.  The solid red
  lines give the  mean relation of the model  galaxies from unweighted
  least-squares fits  of $V$ on  $L$ and $R$  on $L$.  The  colors and
  point    types   correspond    to   surface    brightness    as   in
  Fig.\ref{fig:VLR_ALL}.  Each model  sample consists of $\simeq 1300$
  galaxies,  sampled   to  reproduce  the   observed  distribution  of
  luminosities.  }
\label{fig:VLR6-scatter-BFs}
\end{center}
\end{figure*}

\subsubsection{Bulge Formation}
\label{sec:bulgeeffect}

Thus far we have only considered models without bulge formation (i.e.,
with  $\beta_{\rm   crit}=1$  in  Eq.~(\ref{eqn:betamax})).    We  now
introduce bulge formation to the three models described above.  We set
$\fx=0.25$, and  tune $\beta_{\rm crit}$  so that the sample  of model
galaxies has  the same mean  bulge-to-disk ratio as the  data ($\simeq
0.15$). None of our results are  sensitive to the exact value of $\fx$
chosen.

Since $\mgal$ increases and $c$ decreases with halo mass, more massive
halos typically host galaxies  with a larger $\beta_{\rm max}$.  Also,
for  a  given halo  mass,  the  disk's  contribution to  the  circular
velocity   increases  with  decreasing   spin  parameter   $\lamgal$.  
Consequently,   it  is  predominantly   the  luminous,   high  surface
brightness galaxies residing  in massive halos that will  form a bulge
in our models.

Bulge formation only has a small effect on the $VL$ relation, but does
cause a modest change of the  $RL$ relation. For our fiducial value of
$\fx$, and  in fact for  all $\fx \lta  1$, bulge formation  causes an
increases of the {\it specific} angular momentum of the disk, and thus
of its scale length.  Since bulge formation preferentially affects the
galaxies in the  lower right part of the $RL$  plane (i.e. the highest
surface brightness galaxies), this will actually reduce the scatter in
the $RL$ relation at the bright end (see also Shen \etal 2003). At the
faint end no bulges form, so that this also causes a small increase in
the  slope  of the  $RL$  relation,  which  we counter  by  increasing
$\alpham$.  For realistic bulge-to-disk  ratios the overall scatter in
the $RL$ relation is reduced  at most by $\simeq 15\%$.  Although this
also changes the  maximum amounts of scatter allowed by  the data by a
similar fraction, the maximum scatter in $\lamgal$ allowed by the $RL$
scatter is never  larger than $0.3$, which is  still much smaller than
the predicted amount.   Thus bulge formation is not  able to reconcile
the expected scatter in $\lambda$ with that of the observed scatter in
the $RL$ relation.

The $VL$  and $RL$  relations for models  I-III with  bulge formation,
scatter in  $c$, $\lamgal$, $\YI$, and observational  errors are shown
in Fig.~\ref{fig:VLR6-scatter-BFs}.  Overall, all three models provide
a  reasonable match  to the  slopes, zero  points and  scatter  of the
observed      relations.      A      detailed      comparison     with
Fig.~\ref{fig:VLR_ALL},  however, reveals that  all three  models, but
especially Model~II,  have a  slight problem at  the bright  end where
they predict rotation velocities that  are somewhat too high.  This is
related to the fact that  the contribution of the baryons to $V_{2.2}$
increases  with luminosity (see  \S\ref{sec:massfrac}).  Consequently,
the $VL$ relation  deviates from a pure power-law.  As  we will see in
\S\ref{sec:rescorr}  below,  this   causes  a  weak,  but  significant
correlation between surface brightness and the $VL$ residual.
\begin{figure*}
\begin{center}
\includegraphics[width=6.0in]{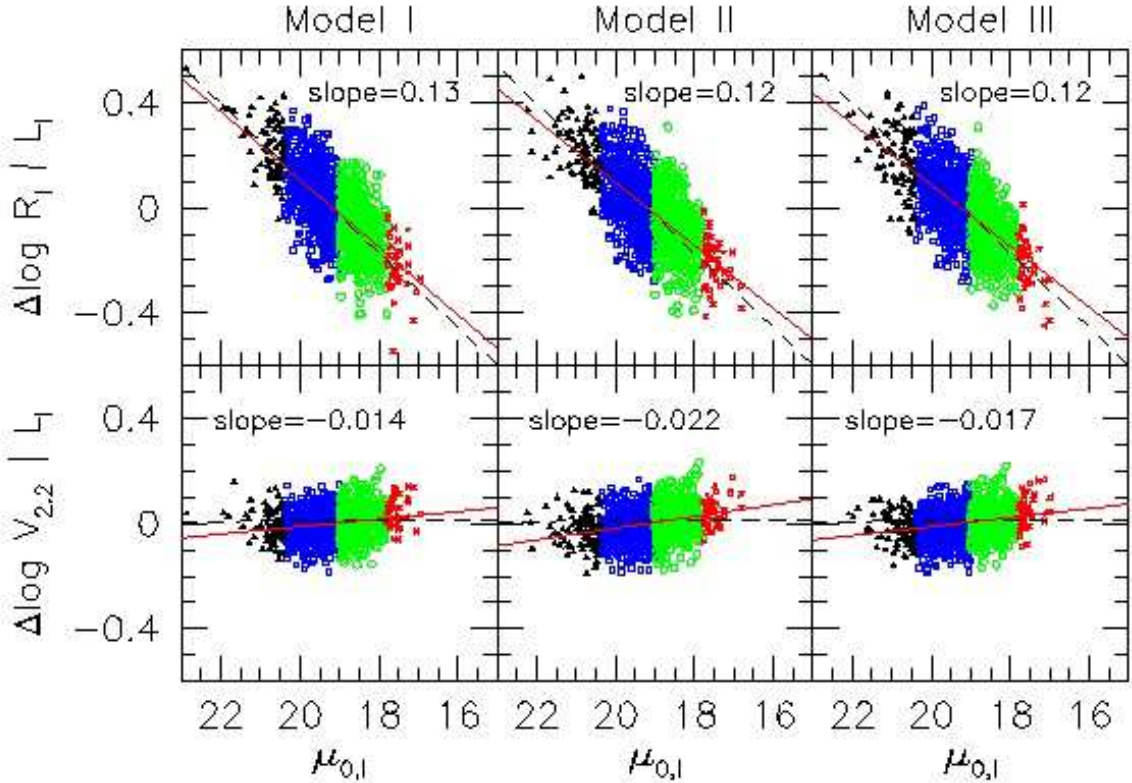}
\caption{Residuals of the model $I$-band $VL$ and $RL$ relations versus surface
  brightness,  for  models  in  Fig.~\ref{fig:VLR6-scatter-BFs}.   The
  observed relations are  given by the dashed line.   All three models
  reproduce the  surface brightness  dependence of the  $RL$ relation,
  but none  of the models are  able to reproduce  the observed surface
  brightness independence of the $VL$ relation.}
\label{fig:dmu6-scatter-BFs}
\end{center}
\end{figure*}

\begin{figure*}
\begin{center}
\includegraphics[width=5.5in]{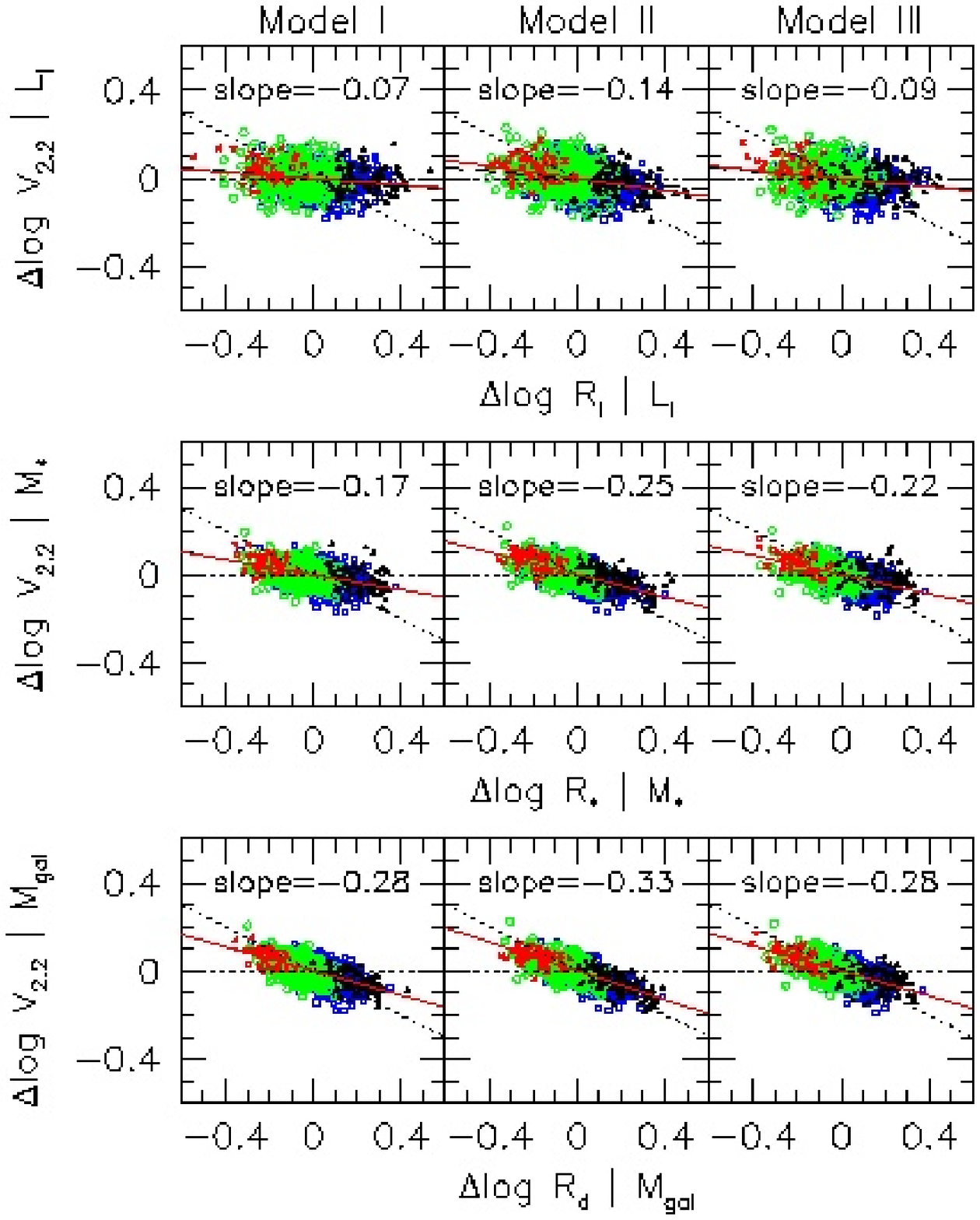}
\caption{Residual correlations for our three models. The upper panels show
  the residuals  of the $I$-band  $VL$ and $RL$ relations,  the middle
  panels  show  the  residuals  of  the stellar  mass  $VM$  and  $RM$
  relations, while the lower panels show the residuals of the baryonic
  mass $VM$ and $RM$ relations.  The  solid red lines show the fits of
  $\Delta V$  on $\Delta  R$. The  dotted lines show  slopes of  0 and
  -0.5.  For  the upper  panels we  show the  observed  correlation as
  long-dashed  lines.    Notice  that  the   correlation  between  the
  residuals  decreases in magnitude  going from  the baryonic  mass to
  stellar mass to I-band luminosity.}
\label{fig:dVLR6-scatter-BFs}
\end{center}
\end{figure*}

\subsection{Residual correlations}
\label{sec:rescorr}

Models I, II  and III all fit the slopes, zero  points and the amounts
of scatter of  the $VL$ and $RL$ relations.   An important question is
whether  we can discriminate  between these  three models,  or whether
there are genuine degeneracies in the model parameter space.  As shown
in  Fig~\ref{fig:dVLR_ALL}  there  is  additional information  in  the
residual correlations.   In particular, the residuals in  the $VL$ and
$RL$ relations are virtually  uncorrelated, so that the $VL$ residuals
are not  significantly correlated with surface  brightness (CR99).  We
now  investigate how  models  I, II  and  III fare  in matching  these
aspects of the data.

The upper panels of Fig.~\ref{fig:dmu6-scatter-BFs} show the residuals
of the $RL$ relations of models I--III plotted against central surface
brightness.   All three  models are  virtually  indistinguishable, and
accurately    reproduce     the    data    (upper-left     panel    of
Fig.~\ref{fig:dVLR_ALL}).    This,  however,   owes   simply  to   the
luminosity  sampling   of  the  data,  which  we   reproduce  using  a
Monte-Carlo technique (see  \S\ref{sec:datafit}).  The lower panels of
Fig.~\ref{fig:dmu6-scatter-BFs},  however,  plot  the  $VL$  residuals
against  central  surface brightness.   This  time,  the three  models
differ slightly from each other,  and significantly from the data. All
models predict a small  but non-negligible correlation between surface
brightness  and the $VL$  residual, contrary  to the  data (lower-left
panel  of  Fig.~\ref{fig:dVLR_ALL}).   This  owes mainly  to  the  non
power-law character of the model $VL$ relations, as discussed above.

The  upper panels  of Fig.~\ref{fig:dVLR6-scatter-BFs}  show  the $VL$
residuals  as function of  the $RL$  residuals.  Except  for Model~II,
which  predicts a  correlation  slope $\gamma  \equiv \dd[\Delta  \log
V(L)]  / \dd[\Delta  \log R(L)]$  which is  somewhat too  steep, these
residual correlations are in good agreement with the data (lower-right
panel of Fig.~\ref{fig:dVLR_ALL}).

As  discussed  in  \S\ref{sec:intro},  the  absence  of  a  pronounced
correlation between the  $VL$ and $RL$ residuals is  contrary to naive
expectations. In order to understand  which aspects of the model cause
this success,  we start by  considering the model {\it  baryonic} $VM$
and $RM$ relations (i.e.  velocity  at 2.2 baryonic scale lengths {\it
  vs.}  baryonic  mass and baryonic  scale length {\it  vs.}  baryonic
mass).  The lower panels of Fig.~\ref{fig:dVLR6-scatter-BFs} show that
the  residuals  of  these  baryonic  scaling  relations  are  strongly
correlated, for each of the three models. This owes to the fact that a
smaller $\lamgal$  implies both a  reduction of the disk  scale length
and an  increase in $V_{2.2}$.   Thus, the naive expectation  that the
residuals should be correlated at least holds for the baryonic scaling
relations.   We find  this  to be  a  very generic  prediction of  our
models,  even though  the exact  value of  the slope  of  the baryonic
residual correlation  may vary somewhat from model  to model. Although
measuring  baryonic  masses  and  baryonic sizes  depends  on  stellar
mass-to-light ratios, which are difficult  to constrain on a galaxy by
galaxy basis,  these can  be determined in  a statistical sense  for a
suitably  chosen  sample.  Thus,  in  principle  the baryonic  scaling
relations are observable, and our predictions regarding their residual
correlations  provide  a useful  test  for  our  models.  We  caution,
however, that it is crucial  that the rotation velocity is measured at
2.2 disk  scale lengths. For example,  when $V$ is measured  at 5 disk
scale lengths,  our models predict a  significantly shallower residual
correlation.

The  middle   panels  of  Fig.~\ref{fig:dVLR6-scatter-BFs}   show  the
residual correlations from the model  {\it stellar} mass $VM$ and $RM$
relations  (i.e.  velocity  at 2.2  stellar  scale lengths  {\it vs.}  
stellar mass and stellar scale  length {\it vs.}  stellar mass).  Note
that the strength of  these correlations is significantly reduced with
respect to  the baryonic case (lower  panels).  This is  caused by the
star formation threshold, which correlates the stellar mass at a given
baryonic  mass to  the  spin  parameter: a  lower  $\lamgal$ not  only
results  in a smaller  scale length  and a  larger $V_{2.2}$,  it also
results in  a relatively larger  stellar mass, therewith  reducing the
residual correlation.  This, together  with the constraint of matching
the  slope of  the  $VL$ relation,  is  another argument  in favor  of
including a star formation  threshold (or something equivalent) in our
models.  In fact, FA00 argued  that a star formation threshold density
is the  main explanation for  the weak residual correlation  observed. 
Although  we   agree  that  it  substantially   reduces  the  residual
correlations,  it  does   not  automatically  result  in  uncorrelated
residuals,    as   is    evident   from    the   middle    panels   of
Fig.~\ref{fig:dVLR6-scatter-BFs}.  The  weaker correlation between the
$VL$ and $RL$  residuals in the upper panels  compared with the middle
panels in Fig.~\ref{fig:dVLR6-scatter-BFs} is achieved through scatter
in  $\YI$  and,   to  a  lesser  extent,  to   our  modelling  of  the
observational errors in $V$, $L$ and $R$.

\subsubsection{Impact of Scatter}
\label{sec:cres}

\begin{figure*}
\begin{center}
\includegraphics[width=6.0in]{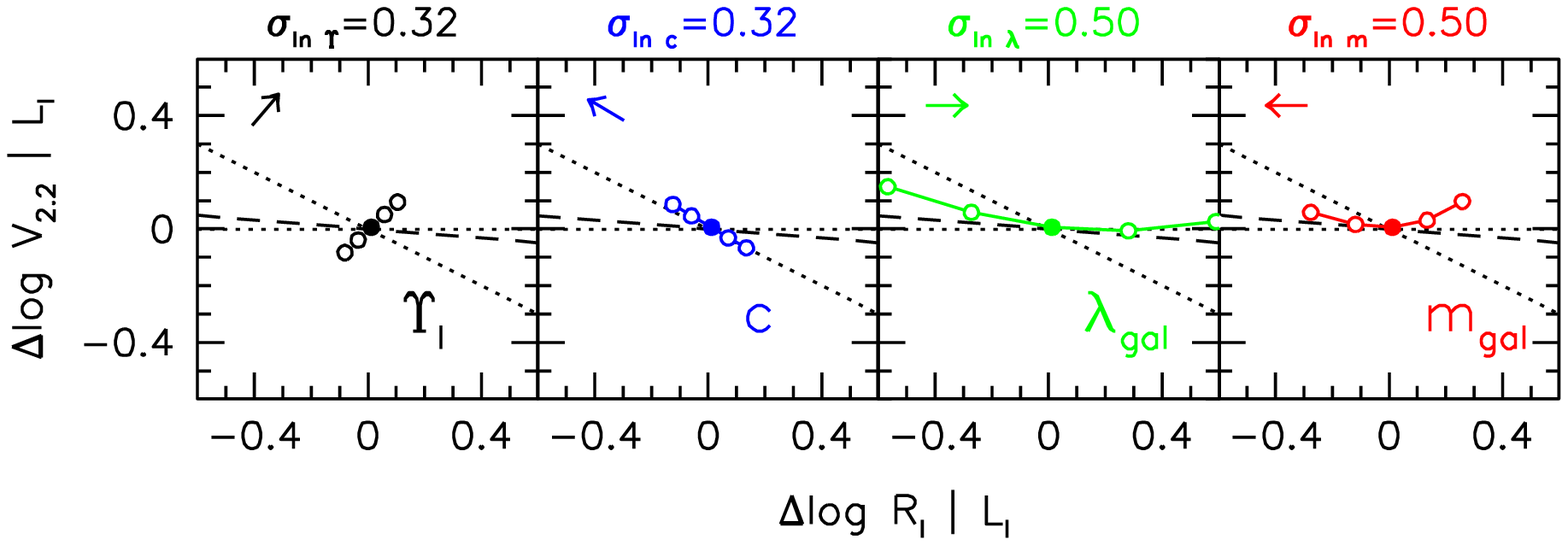}
\caption{Contribution of scatter in $\YI$, $c$, $\lamgal$, and
  $\mgal$  to the residual  correlation for  our reference  model with
  $\Mvir =  3\times 10^{11}\,h^{-1}\,\Msun$ and  $\DeltaIMF=-0.5$ (see
  also  Fig.~\ref{fig:VLR-mcl8}).   The   solid  dot  shows  the  mean
  parameter model,  the open circles show  the models with  $\pm 1$ \&
  $2\sigma$ of the scatter given at  the top of each panel.  The arrow
  in  the top left  of each  panel shows  the direction  of increasing
  parameter.   The observed  slope is  shown by  the dashed  line; the
  dotted lines have slopes of 0  and -0.5. A slope of -0.5 is expected
  for a pure exponential disk (CR99).}
\label{fig:dVLR-mcl}
\end{center}
\end{figure*}

\begin{figure*}
\begin{center}
\includegraphics[width=6.0in]{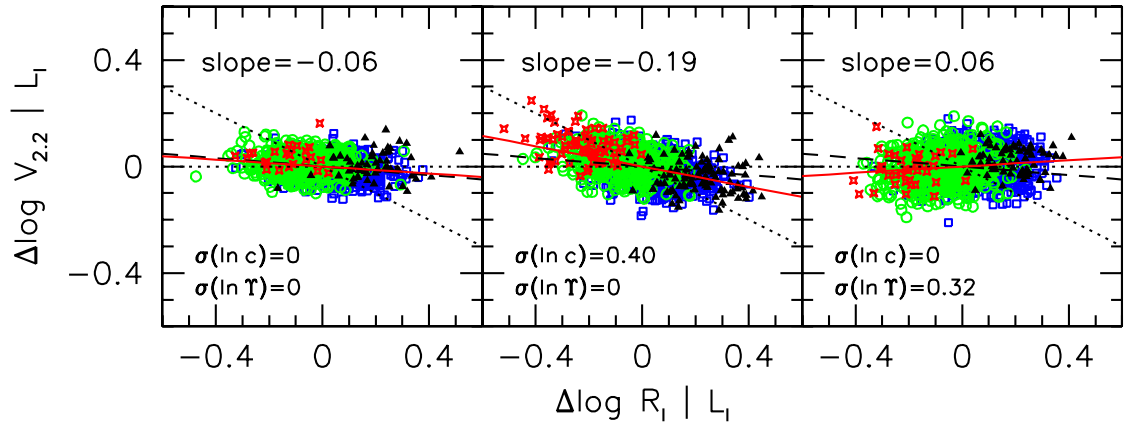}
\caption{Residual correlations for Model I with different amounts of
  scatter   in  $\YI$  and   $c$.   All   panels  are   computed  with
  $\sigma_{\ln\lambda}=0.25$, and observational  errors.  The model on
  the left  has zero  scatter in  $c$ and $\YI$,  and a  weak negative
  residual  correlation consistent with  the data.   The model  in the
  middle has  maximum scatter  in $c$, this  results in  a significant
  negative residual  correlation. The model  on the right  has maximum
  scatter  in  $\YI$,  this  results  in a  slight  positive  residual
  correlation. Thus larger  scatter in $\YI$ and lower  scatter in $c$
  help to reduce the magnitude of the residual correlation.  }
\label{fig:dVLR3}
\end{center}
\end{figure*}

To  gauge how  the  various  sources of  scatter  impact the  residual
correlations,  Fig.\ref{fig:dVLR-mcl}  shows  $\Delta\log  V$  against
$\Delta\log R$  for models that have  only scatter in one  of the four
parameters: $\YI$,  $c$, $\lamgal$,  or $\mgal$.  The  spin parameter,
$\lamgal$,  and   the  galaxy  mass  fraction,   $\mgal$,  only  cause
significant scatter  in the $RL$ relation.   Consequently, they result
in  residual  correlations with  $\vert  \gamma  \vert  \lta 0.1$,  in
agreement  with the data,  which has  $\gamma =  -0.08 \pm  0.03$ (see
\S\ref{sec:datafit}).   Scatter  in $c$  and  $\YI$, however,  predict
strongly    correlated   residuals    with    $\gamma\simeq-0.6$   and
$\gamma\simeq1$, respectively.

Thus, achieving  $VL$ and  $RL$ relations with  uncorrelated residuals
requires a subtle  balance between the amounts of  scatter in the halo
concentrations, $c$, and the  stellar mass-to-light ratios, $\YI$.  To
emphasize   this  point,   Fig.\ref{fig:dVLR3}   shows  the   residual
correlations for models  with different amounts of scatter  in $c$ and
$\YI$.  The scatter in $\lamgal$  and the scatter due to observational
errors are kept constant.  A  model with only scatter in $\lamgal$ and
observational errors has $\gamma=-0.06$.  Including the maximum amount
of  scatter  in  $c$  (middle  panel) results  in  a  strong  negative
correlation,  $\gamma=-0.19$.   In  contrast,  including  the  maximum
scatter in $\YI$, with no scatter  in $c$, results in a weak, positive
correlation  with  $\gamma\simeq0.06$   (right  panel).   In  practice
scatter in both  $c$ and $\YI$ are expected to  contribute to the $VL$
scatter,  so  there   is  not  much  freedom  to   tune  the  residual
correlations. In order to reproduce  the total $VL$ scatter we adopted
$\sigma_{\ln     c}=\sigma_{\ln\Upsilon}=0.23$.     As     shown    in
Figs.~\ref{fig:dmu6-scatter-BFs}  \& \ref{fig:dVLR6-scatter-BFs}, this
amount of scatter  results in a weak residual  correlation and a small
but  significant surface brightness  dependence to  the $VL$  scatter. 
Removing these dependences entirely  would require a larger scatter in
$\YI$ and  a lower  scatter in  $c$. The former  is feasible,  but the
latter would  require an  even more biased  subset of halos  than just
those without recent major mergers.

\begin{figure*}
\begin{center}
\includegraphics[width=6.0in]{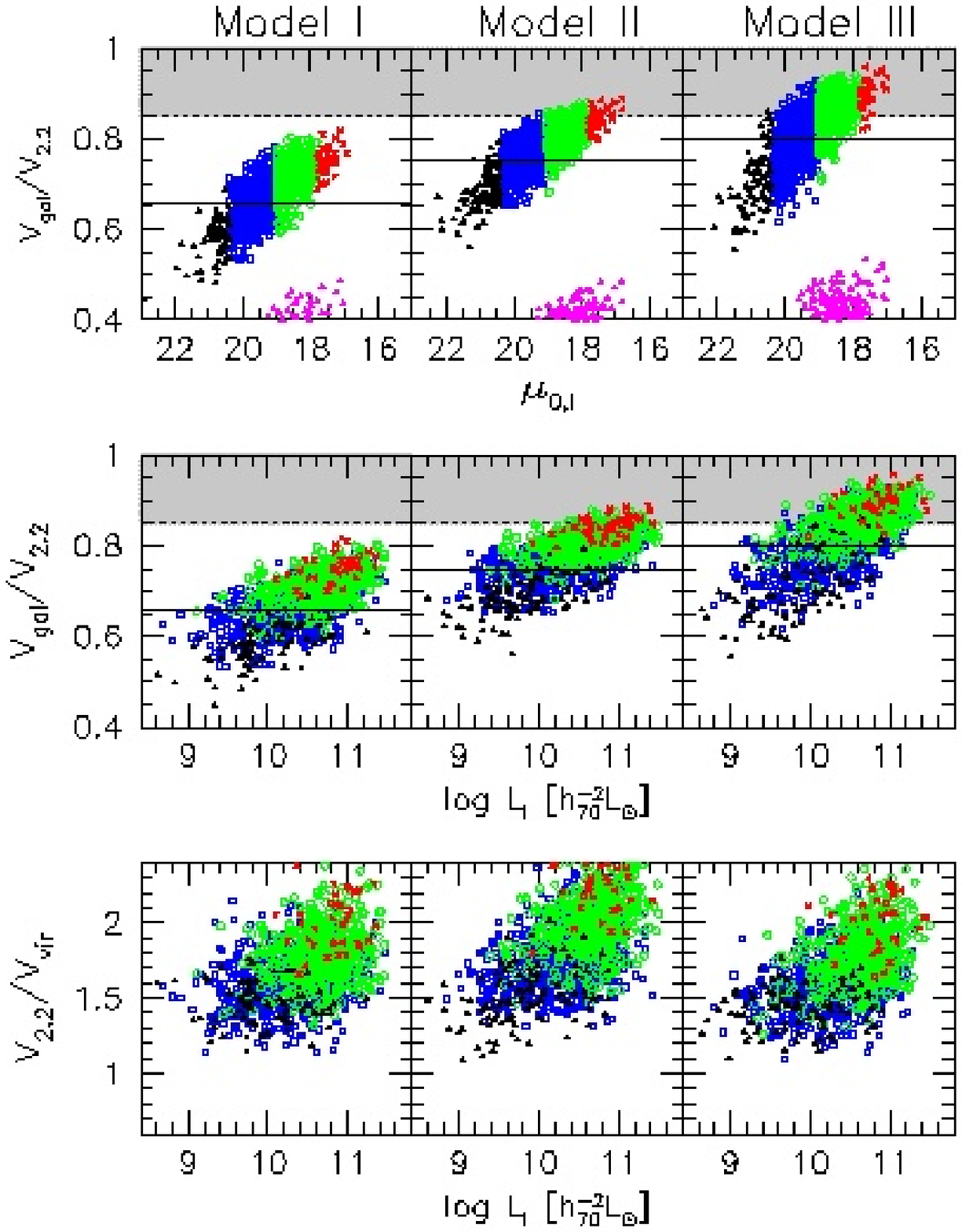}
\caption{Contribution of the baryons (stellar disk, gaseous disk, and
  bulge)  to circular  velocity  at 2.2  disk  scale lengths,  $V_{\rm
    gal}/V_{2.2}$, and  ratio of  observed to virial  circular velocity
  for Models I, II, \& III.  Point types and colors are the same as in
  Fig.~\ref{fig:VLR_ALL}.   All  models show  a  strong dependence  of
  $V_{\rm gal}/V_{2.2}$ with surface  brightness (upper panels), and a
  weaker  one  with luminosity  (middle  panels).   The dependence  on
  luminosity  is  mostly because  more  luminous  galaxies have  higher
  surface  brightness.   The  horizontal  dotted  line  shows  $V_{\rm
    gal}/V_{2.2}=0.85$, the shaded  region above this line corresponds
  to galaxies that are baryon dominated within 2.2 disk scale lengths.
  Note however that due to our bulge formation recipe the contribution
  of the disk  to $V_{\rm gal}$ is always less  than 0.85. The maximum
  contribution of the disk to $V_{\rm gal}$ is given by the horizontal
  black solid  line.  The  contribution of the  bulge is shown  by the
  magenta  triangles,   and  in   all  cases  contributes   less  than
  $\simeq25\%$ of the  total mass within 2.2 disk  scale lengths.  The
  lower  panels show  $V_{2.2}/\Vvir$ against  luminosity.  The median
  $V_{2.2}/\Vvir$ for  these model  I, II, \&  III are 1.68,  1.86, \&
  1.67 respectively.  }
\label{fig:vdt6-scatter-BFs}
\end{center}
\end{figure*}

\subsubsection{Dark and baryonic mass fractions}
\label{sec:massfrac}

The  upper panels  of Fig.~\ref{fig:vdt6-scatter-BFs}  show  the ratio
$V_{\rm gal}/V_{2.2}$ as function  of surface brightness for all three
models.  Here $V_{\rm gal}$  and $V_{2.2}$ are the circular velocities
of  the baryons  (disk plus  bulge)  and the  total mass  distribution
(baryons  plus dark matter)  respectively, both  measured at  2.2 disk
scale  lengths.   The  solid,  horizontal line  indicates  $\beta_{\rm
  crit}$:  all galaxies  above  this line  contain  a bulge  component
(whose contribution to  $V_{2.2}$ is given by the  magenta triangles). 
Since  we tuned  $\beta_{\rm crit}$  to match  the  mean bulge-to-disk
ratio of the data, its value is different for different models.

In  all 3 models,  $V_{\rm gal}/V_{2.2}$  is strongly  correlated with
surface brightness, and to  a lesser extent luminosity (middle panels;
see also FA00 and Zavala \etal 2003).  In fact, most of the luminosity
dependence  simply  owes  to  the  fact that  more  luminous  galaxies
typically have  a higher  surface brightness, as  is evident  from the
color-coding.    Note   that  the   mean   $V_{\rm  gal}/V_{2.2}$   is
significantly different for different  models.  In the case of Model~I
(top-heavy  IMF), there are  no galaxies  with $V_{\rm  gal}/V_{2.2} >
0.85$ (indicated by the gray shaded region).  This means that, in this
model, the baryons never contribute more than $\sim 70$ percent of the
total enclosed mass within $2.2\Rd$. In models II and~III the fraction
of  galaxies with  $V_{\rm gal}/V_{2.2}  > 0.85$  are  $\simeq4\%$ and
$\simeq37\%$, respectively. Note that Models~I and~III, though, have a
very  similar residual  correlation  slope $\gamma$  (upper panels  of
Fig.~\ref{fig:dVLR6-scatter-BFs}).  This  is in contradiction  to CR99
who argued that $\gamma$ is  uniquely correlated with the mean $\vdt$;
in  particular,  from the  weak  residual  correlation observed,  they
concluded that high surface brightness disk galaxies should have, {\it
  on average} $\vdt\simeq 0.6$.  In Appendix~\ref{sec:CR99} we discuss
the CR99  method in more  detail, and show  why the observed  value of
$\gamma$ does not accurately constrain $\vdt$.

Finally, the lower  panels of Fig.~\ref{fig:vdt6-scatter-BFs} show the
ratio  $V_{2.2}/\Vvir$ as  a function  of luminosity.   Note  that the
circular velocity  at 2.2 disk  scale lengths can  be as large  as two
times the  virial velocity. The  median values of  $V_{2.2}/\Vvir$ are
$1.68$, $1.86$ and $1.67$ for Models~I, II, and~III, respectively.  As
we  discuss  in  more   detail  in  \S\ref{sec:revised}  below,  these
relatively high values have important implications.


\section{Low spin parameter Models}
\label{sec:lowspin}

As discussed the zero point  of the $RL$ relation is highly degenerate
between  changes in $\overline{\lambda}_{\rm  gal}$ and  $\mgalo$.  In
\S\ref{sec:convert} we  chose to  tune $\mgal$ in  order to  match the
$RL$ relation,  while keeping $\overline{\lambda}_{\rm  gal}$ fixed at
the  median  value  of  the  spin  parameter  of  dark  matter  halos;
$\overline{\lambda}_{\rm gal} = \overline{\lambda} = 0.042$.
%
However, there are several reasons  why one might expect that the spin
parameter  of the galaxy  is different  from that  of its  dark matter
halo.   Firstly, numerical  simulations suggest  that haloes  that are
more likely to host disk galaxies, i.e., those that did not experience
any  recent  major mergers,  have  a  median  spin parameter  that  is
significantly  lower than  that  of the  full  distribution of  haloes
(D'Onghia \& Burkert 2004;  Macci{\`o} \etal 2006).  Secondly, mass is
generally more centrally concentrated  than specific angular momentum. 
Consequently,  if disks  form inside  out, one  expects that  $\jgal <
\mgal$,  and  thus  $\lamgal   <  \lambda$.   Thirdly,  during  galaxy
formation angular momentum may be  transferred from the baryons to the
halo via  dynamical friction, also  resulting in $\lamgal <  \lambda$. 
Finally, numerical simulations have shown that the {\it directions} of
the angular  momentum vectors  of the (hot  and cold) baryons  and the
dark matter can be strongly misaligned (van den Bosch 2002; Chen \etal
2003; Sharma  \& Steinmetz 2005;  Bailin \etal 2005b),  suggesting that
$\lambda$ and $\lamgal$ are only poorly correlated.

\begin{figure*}
\begin{center}
\includegraphics[width=6.0in]{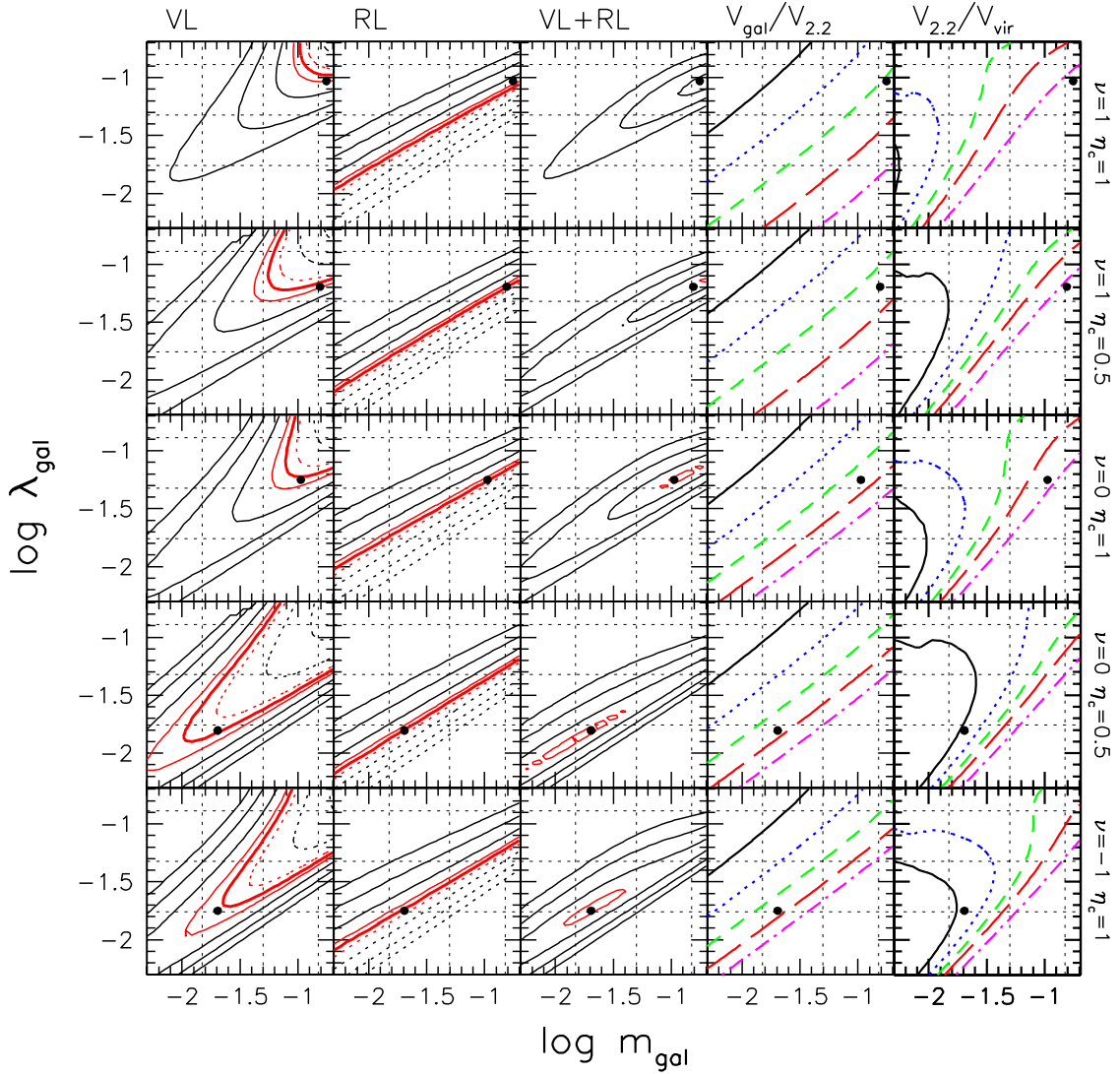}
\caption{
  Parameter space of models for $0.005 \le \mgal,\lamgal \le 0.2$. All
  models  assume $\DeltaIMF=-0.2$  and have  no bulge  formation.  The
  horizontal lines  show $\lamgal=0.018, 0.048$, \&  $0.129$ (the mean
  and 2$\sigma$  scatter in $\lambda$ for cosmological  DM halos.  The
  vertical   dotted  lines   show  $\mgal=0.015,   0.05$,   \&  $0.15$
  corresponding  to $\simeq  10\%, 33\%$  \& $100\%$  galaxy formation
  efficiency.   Each row  corresponds  to a  different combination  of
  adiabatic  contraction,   $\nu$,  and  concentration  normalization,
  $\eta_c$.  The  first and second  columns show the offsets  from the
  $VL$  and $RL$  relations  at $\log\LI=10.3$,  with  respect to  the
  observed scatter in these  relations.  The black contours correspond
  to 1,2,\&  3 times the observed  scatter. The red  contours show 0.3
  times  the observed  scatter,  which is  approximately the  observed
  uncertainty in the  zero points.  The thick red  line corresponds to
  no offset.   Positive offsets are contoured with  solid lines, while
  negative offsets are contoured  with dotted lines.  The third column
  shows the  quadratic sum  of the $VL$  and $RL$ offsets.   The solid
  black dot  shows the minimum of  the $VL+RL$ offset  within our grid
  for each model.   The fourth column gives the  ratio of the circular
  velocity  of the disk  to the  total circular  velocity at  2.2 disk
  scale lengths, $V_{\rm d}/V_{2.2}$: 0.95 (magenta dot-short dashed);
  0.87  (red  long  dashed);  0.71  (green short  dashed);  0.5  (blue
  dotted); 0.3 (black solid). The  fifth column gives the ratio of the
  circular velocity at 2.2 disk scale lengths to the circular velocity
  at  the  virial  radius,  $V_{2.2}/\Vvir$:  2.0  (magenta  dot-short
  dashed); 1.6 (red long dashed);  1.4 (green short dashed); 1.2 (blue
  dotted);  1.0 (black  solid.  The  top row  uses  standard adiabatic
  contraction, $\nu=1$, and  concentration parameters, $\eta_c=1$: The
  second   row   also  uses   standard   adiabatic  contraction,   but
  concentration parameters a factor 2 lower than the Bullock \etal (2001a)
  model, $\eta_c=0.5$.  The third  row shows a model without adiabatic
  contraction, $\nu=0$,  but standard concentrations,  $\eta_c=1$.  In
  order to match the zero points with low $\lamgal$, we need to either
  reduce  the  concentration,  $\eta_c=0.5$  (fourth  row),  or  have
  expansion of the halo, $\nu=-1$ (fifth  row).}
\label{fig:chi5}
\end{center}
\end{figure*}

To  fully explore these  possibilities, we  relax the  constraint that
$\overline{\lambda}_{\rm  gal}  =   \overline{\lambda}  =  0.042$  and
investigate which combinations of $\lamgal$ and $\mgal$ match the zero
points of the $VL$ and $RL$  relations.  Our strategy is to run a grid
of models in $\mgal$-$\lamgal$  space while keeping all other parameters
fixed.    In  particular,  we   set  $Q=1.5$,   $\DeltaIMF=-0.2$,  and
$\beta_{\rm crit}=1$.   For a  given combination of  the concentration
normalization,  $\eta_c$,  and  the adiabatic  contraction  parameter,
$\nu$, we  compute the zero point  offsets from the  observed $VL$ and
$RL$ relations.  We set the zero point to $\log\LI=10.3$.  Rather than
compute  a full  model for  each $\lamgal$  and $\mgal$,  we  find the
values  of $\Mvir$ that  result in  $\log\LI=10.3$.  This  requires an
iterative  procedure since the  stellar mass  fraction, and  hence the
luminosity,  depend   on  $\lamgal$.   In  order   to  facilitate  the
interpretation  of the  zero point  offsets we  normalize them  to the
observed scatter  in the $VL$  and $RL$ relations,  respectively.  The
results of these grid  searches are shown in Fig.\ref{fig:chi5}.  Each
row corresponds  to a  model with a  different $\eta_c$ and  $\nu$, as
indicated.

Panels  in  the  first and  second  columns  from  the left  show  the
constraints on $\lamgal$ and $\mgal$ from the $VL$ zero points and the
$RL$  zero point,  respectively. For  the  $RL$ relation,  there is  a
strong  degeneracy  in  all models  with  $\lamgal\propto\mgal^{0.5}$;
however,  for a  given $\mgal$  the  constraint on  $\lamgal$ is  very
strong. The third column from the  left shows the quadratic sum of the
$VL$ and $RL$ offsets, with contours  as in the first two columns. The
solid black  dot shows the  minimum of the  $VL + RL$ offset  for each
model within our grid, so it  is not necessarily a global minimum. The
fourth column from  the left shows the ratio  of the circular velocity
of  the baryons  to  the total  circular  velocity at  2.2 disk  scale
lengths,   $V_{\rm   gal}/V_{2.2}$.    Lines   of   constant   $V_{\rm
  gal}/V_{2.2}$  are  approximately  $\lamgal\propto\mgal$.  In  these
models  $V_{\rm   gal}/V_{2.2}$  is  equivalent   to  central  surface
brightness, so  that galaxies with  low surface brightness  disks have
low $V_{\rm gal}/V_{2.2}$, and  vice versa for high surface brightness
disks.  The fifth column from the left shows the ratio of $V_{2.2}$ to
$\Vvir$. The dependence of this ratio on $\mgal$ and $\lamgal$ is more
complicated than $V_{\rm gal}/V_{2.2}$ but follows the same trend that
models   with  higher   $\mgal$  and   lower  $\lamgal$   have  higher
$V_{2.2}/\Vvir$.  Thus for  model galaxies  that match  the  $RL$ zero
point, lower $\lamgal$ solutions  have lower $V_{\rm gal}/V_{2.2}$ and
$V_{2.2}/\Vvir$. We discuss the significance of this result below.

The first row uses the standard halo concentrations ($\eta_c=1.0$) for
a \LCDM  concordance cosmology ($\OmegaM=0.3$, $\Omega_{\Lambda}=0.7$,
$\sigma_8=0.9$) and standard  adiabatic contraction ($\nu=1.0$).  Note
that   the    standard   model    advocated   in   MMW,    which   has
$\mgal=\lamgal=0.05$, predicts  a $VL$ zero point that  is offset from
the data  by $\sim 2\sigma$.  Matching  the $VL$ zero  point for these
values of  $\eta_c$ and $\nu$  requires $\mgal\simeq\lamgal\simeq0.1$. 
However,  this clearly  results in  disks  that are  too large.   This
demonstrates that  this `standard'  model is unable  to simultaneously
match the $VL$ and $RL$ zero points for realistic $\YI$.

The  second row  shows  a  model with  $\eta_c=0.5$,  i.e.  with  halo
concentrations  that are  50\% lower  than for  the  \LCDM concordance
cosmology.   Although one  can  now  match the  $VL$  zero point  with
smaller, more  realistic $\lamgal$, a  simultaneous match to  the $VL$
and $RL$ zero points still  requires relatively high $\mgal$.  This is
essentially   Model~II.     As   shown    in   the   third    row   of
Fig.\ref{fig:chi5}, very similar results  are obtained for models with
standard  halo  concentrations   ($\eta_c=1$)  but  without  adiabatic
contraction ($\nu=0$).   Model~III is basically  an example of  such a
model, which indeed  yields results that are very  similar to those of
Model~II.

In  the fourth  row  we consider  an  even more  unorthodox model;  in
addition to `turning off' adiabatic contraction ($\nu=0$), we also set
the halo  concentrations to be 50\%  lower than expected  in the \LCDM
concordance cosmology ($\eta_c=0.5$).   For these parameters, the $VL$
zero  point can  be matched  for a  very wide  range of  $\lamgal$ and
$\mgal$.  In  fact, there are two branches  in $\mgal$-$\lamgal$ space
that match the $VL$ zero  point, corresponding to high and low surface
brightness disks.   This branching is also visible  from the curvature
of  the $\lamgal$  and  $\mgal$ lines  in  the upper  right panels  of
Fig.~\ref{fig:VLR-mcl8}. In order to  also match the $RL$ relation the
model needs to  fall on the high surface  brightness branch, and needs
to have values for both $\lamgal$ and $\mgal$ that are much lower than
for the previous models.  Similar constraints are obtained for a model
with halo  expansion ($\nu=-1$) but with  standard halo concentrations
($\eta_c=1$), shown in the fifth row.

\begin{figure*}
\begin{center}
\includegraphics[width=6.0in]{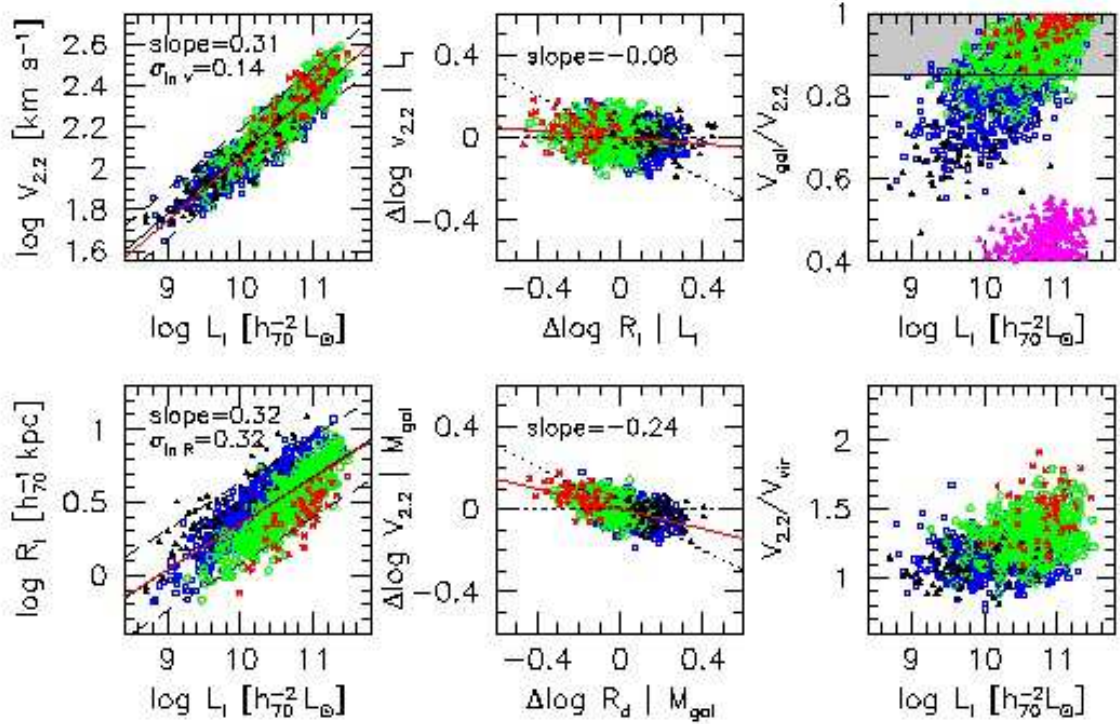}
\caption{Scaling relations and residual correlations of model IV. This
  model reproduces the slopes, scatter and zero points of the $VL$ and
  $RL$ relations  as well as  the weak residual correlation.   As with
  models I,  II, \& III, model  IV predicts that the  residuals of the
  baryonic   mass  $VM$   and  $RM$   relations  should   be  strongly
  anti-correlated.  This  model contains  both baryon and  dark matter
  dominated  galaxies at  2.2 disk  scale  lengths, and  has a  median
  $V_{2.2}/\Vvir=1.22$   which   is  often   required   in  order   to
  simultaneously fit the luminosity function and the $VL$ relation.  }
\label{fig:MV6}
\end{center}
\end{figure*}

The lower  two rows of  panels thus indicate  that there is a  part of
parameter space that  can match the $VL$ and  $RL$ zero points equally
well as models I-III, but with a much lower average spin parameter and
much  lower galaxy  mass  fractions. Before  we  address the  physical
relevance  of  these unconventional  models,  we  investigate to  what
extent  they can match  the $VL$  and $RL$  slopes and  their residual
correlations.  Fig.~\ref{fig:MV6} shows scaling relations and residual
correlations for  one particular model (hereafter  Model~IV) with halo
expansion and standard concentrations, which matches both the $VL$ and
$RL$ zero points.  The parameters of this model are listed in Table~2,
for  the  scatter  in  the  model  parameters  we  adopt  $\sigma_{\ln
  c}=\sigma_{\ln   \Upsilon}=0.23$   (as   for   models   I-III)   and
$\sigma_{\ln  \lambda}=0.28$.   The median  galaxy  spin parameter  is
$\overline{\lambda}_{\rm gal}=0.023$,  which is similar  to the median
spin parameter of  dark matter halos that have not  had a recent major
merger (D'Onghia  \& Burkert  2004), while $\mgalo  = 0.03$.   Both of
these values  are much  lower than for  models I-III.  Note  that this
model fits  the slopes, zero  points, scatter and  correlation between
residuals  of  the  $VL$  and  $RL$ relations.   It  also  predicts  a
significant  correlation between  the residuals  of  the corresponding
baryonic  scaling  relations, in  agreement  with  Models I-III.   The
average $V_{\rm  gal}/V_{2.2}$ of this  model is even larger  than for
Model~III  ($\simeq  62\%$  percent   of  all  galaxies  have  $V_{\rm
  gal}/V_{2.2}>0.85$), indicating  that a larger  fraction of galaxies
is baryon dominated within $2.2$ disk scale lengths.

Finally,   the   lower-left    panel   of   Fig.~\ref{fig:MV6}   plots
$V_{2.2}/\Vvir$ as  function of  the galaxy $I$-band  luminosity. This
model predicts an average $V_{2.2}/\Vvir$ of $\sim 1.2$, significantly
lower than for models I-III. Thus, $V_{2.2}$ in Model~IV is comparable
to $V_{\rm max}$, the maximum  circular velocity of a NFW halo without
adiabatic contraction (see  Fig.~\ref{fig:Vvir}). This is also evident
from the panels in the fifth column of Fig.\ref{fig:chi5}, which shows
contours  of constant  $V_{2.2}/\Vvir$: note  that the  models  in the
upper  three rows  that simultaneously  match the  $VL$ and  $RL$ zero
points, indicated by  a solid dot, all predict  relatively high values
of $V_{2.2}/\Vvir$ with a mean  of $\simeq 1.8 \pm 0.1$.  However, the
models in  the lower two rows,  which can simultaneously  fit the $VL$
and  $RL$ zero  points with  low $\lamgal$  and $\mgal$,  predict much
lower values of $V_{2.2}/\Vvir$ with a mean of $\simeq 1.2$.

\subsection{A revised model for disk formation}
\label{sec:revised}

We have  shown that  one can construct  models that match  the slopes,
zero points and residual correlations  of the $VL$ and $RL$ relations,
and which predict relatively low values for $V_{2.2}/\Vvir$.  This has
an important  implication: all semi-analytic  models that are  able to
match the  luminosity function (LF)  require $\Vvir \lta  V_{2.2} \lta
V_{\rm  max}$ in  order to  simultaneously match  the $VL$  zero point
(e.g., Somerville \& Primack 1999; Cole \etal 2000; Benson \etal 2003;
Croton \etal 2006).   A similar conclusion was obtained  by Yang \etal
(2003) and van den Bosch \etal (2003) using the conditional luminosity
function formalism.  This suggests that models I-III will be unable to
simultaneously fit  the observed LF  of galaxies.  Model  IV, however,
which is  based on halo expansion, predicts  ratios of $V_{2.2}/\Vvir$
that clearly allow for a simultaneous  fit to the LF.  In more general
terms,   as  can   be  seen   from  the   fifth  row   of   panels  in
Fig.\ref{fig:chi5},  low  values  of $V_{2.2}/\Vvir$  require  average
values for $\lamgal$ and $\mgal$ that are much lower than the standard
values  normally  adopted.   In  the standard  model  (i.e.,  $\nu=1$,
$\eta_c=1$), however, such low values for $\lamgal$ and $\mgal$ result
in a  $VL$ zero-point that is  in violent disagreement with  the data. 
The only models for  which a low $V_{2.2}/\Vvir$ simultaneously allows
a fit to both the $VL$ and $RL$ zero points are those in the lower two
rows: either  adiabatic contraction does not occur  ($\nu=0$) and halo
concentrations are  a factor  two smaller than  predicted for  a \LCDM
concordance cosmology, or halos  actually expand during disk formation
(with  $\nu  \simeq  -1$),   in  which  case  their  original  (before
expansion)  concentrations can be  as predicted.   Of course,  one can
also construct models with intermediate  values of $\nu$ and $\eta_c$. 
For  a cosmology with  $(\OmegaM=0.25,\sigma_8=0.75)$ as  advocated by
the third  year WMAP results (Spergel \etal  2006) $\eta_c=0.75$ which
requires $\nu\simeq -0.5$.

This obviously raises the  question whether halo expansion during disk
formation is a  realistic option from a physical point  of view. It is
certainly inconsistent with the  standard picture, in which disks form
out  of cooling flows  that preserve  their specific  angular momentum
(Fall \&  Efstathiou 1980; MMW;  Dalcanton, Spergel \& Summers  1997). 
Since the  time scale for new gas  to reach the disk  is never shorter
than  the free fall  time, the  formation of  the resulting  disk will
cause  a  contraction  of  the  dark  matter halo  that  is  close  to
adiabatic.  In  order to avoid  contraction, one needs to  abandon the
idea that gas reaches the disk via a (relatively smooth) cooling flow.
Rather, the  disk material  has to assemble  out of  several (massive)
clumps,  which reach the  center of  the halo  by dynamical  friction. 
After  all,  dynamical friction  transfers  the  potential energy  and
orbital angular momentum  of the clumps to the  dark matter particles,
which consequently  move to larger  radii (e.g., El-Zant,  Shlosman \&
Hoffman 2001; Ma \& Boylan-Kolchin  2004; Mo \& Mao 2004; Tonini, Lapi
\& Salucci  2006).  The  same mechanism has  also been proposed  as an
explanation  for the large,  constant density  cores observed  in some
clusters of galaxies (El-Zant \etal 2004; Zappacosta \etal 2006).
 
In  addition  to providing  a  scenario  for  disk formation  that  is
consistent  with the  galaxy luminosity  function, halo  expansion may
also help to alleviate  a long-standing problem with observed rotation
curves, which are  claimed to be inconsistent with  steeply cusped NFW
profiles (e.g., Dutton \etal 2005 and references therein) and with the
observed pattern  speeds of bars  (e.g., Debattista \&  Sellwood 2000,
Weiner \etal  2001).  Although we  defer a detailed discussion  on the
rotation curve shapes of our various models to a forthcoming paper, we
have  verified that  in all  models discussed  here the  galaxies have
realistic (i.e., flat) rotation curves.

Additional `support' for Model~IV  comes from various estimates of the
baryonic  mass  fractions  of  disk  galaxies.   As  is  evident  from
Fig.\ref{fig:chi5}, models with halo expansion that match the $VL$ and
$RL$ zero points  predict much lower values for  $\lamgal$ and $\mgal$
than  the standard  models  with adiabatic  contraction.   As we  have
argued above,  there are  several reasons for  expecting $\lambar_{\rm
  gal} <  \lambar$. Support for  low galaxy mass fractions  comes from
various  sources,   including  the  conditional   luminosity  function
formalism  (Yang  \etal 2003,  2005;  Cooray  2005) and  galaxy-galaxy
lensing (Guzik \& Seljak 2002; Hoekstra \etal 2005). In particular, in
one of  the largest galaxy-galaxy lensing studies  to date, Mandelbaum
\etal (2006) find that late-type galaxies with a stellar mass of a few
times  $10^{10} h^{-1}  \Msun$,  have on  average,  $\mgal \sim  0.03$
(i.e., corresponding to the parameter $\eta$ in Mandelbaum \etal being
$0.18$),  in  good  agreement  with  the value  of  $\mgalo$  for  our
Model~IV.

The main potential problem  for the `expansion-scenario' proposed here
is that it is not clear  whether the merger between various clumps can
produce   a   realistic  disk   galaxy.    Although  high   resolution
hydrodynamical  simulations  of   disk  formation  in  a  cosmological
framework  have recently  suggested that  realistic disk  galaxies may
form  out of merging  progenitors (e.g.,  Robertson \etal  2006), more
work is  clearly needed  to investigate such  a formation  scenario in
detail.

To  summarize,  if  one is  willing  to  abandon  the idea  that  disk
formation  involves adiabatic  contraction of  the  corresponding dark
matter   halo,  one   can   construct  disk   formation  models   that
simultaneously  match  the  $VL$  and $RL$  relations  using  standard
cosmological  parameters.   In   addition,  these  model  predict  low
baryonic mass  fractions, in better  agreement with the  data, predict
low spin  parameters, consistent  with a picture  in which  disks form
preferentially  in halos  with quiescent  merger histories,  and, most
importantly,  yield  characteristic  values for  $V_{2.2}/\Vvir$  that
suggest  that  one  may  be  able to  simultaneously  fit  the  galaxy
luminosity function.


\section{Summary}
\label{sec:summary}

We have  used the slopes,  zero-points, and residuals of  the observed
$VL$  and  $RL$  relations   to  place  constraints  on  the  standard
\LCDM-based  model for  the formation  of disk  galaxies.   Our models
consider  exponential  (baryonic)  disks  in  centrifugal  equilibrium
within NFW dark matter halos.  We  model the reaction of the dark halo
to disk formation in a way which permits it to range from the standard
adiabatic  contraction to an  expansion of  a similar  magnitude.  The
disk stars-to-gas  ratio is determined by a  threshold surface density
for star  formation. A  bulge is included  based on  a self-regulating
mechanism  that  ensures  disk  stability.  The  disk  properties  are
converted   into   observables   using  an   empirically   determined,
luminosity-dependent, stellar mass-to-light ratio.

We construct samples of  model galaxies including intrinsic scatter in
halo  concentrations, stellar  mass-to-light ratios,  and  galaxy spin
parameter.  In  addition, we  mimic observational errors,  and sample
the  model galaxies  so that  they reproduce  the  observed luminosity
distribution.   For  each sample  we  construct  model  $VL$ and  $RL$
relations which  we compare to the  data.  By demanding  the models to
{\it simultaneously} reproduce the  slopes, zero points and scatter of
the  observed  $VL$  and  $RL$ relations,  including  the  correlation
between the residuals, we obtain the following conclusions:

\begin{itemize}
  
\item Since the stellar  mass-to-light ratio increases with increasing
  luminosity, the slopes  of the $VL$ and $RL$  relations deviate from
  the    basic   halo   virial    relation   $\Vvir\propto\Rvir\propto
  \Mvir^{1/3}$.   The  observed  $VL$  slope is  reproduced  when  the
  gas-to-stellar mass ratio is  properly decreasing with mass. This is
  naturally   achieved  by  a   surface-density  threshold   for  star
  formation.   The observed  $RL$ slope  requires that  the  disk mass
  fraction is properly increasing with mass, in agreement with generic
  predictions of  supernova feedback (e.g., Dekel \&  Silk 1986; Dekel
  \& Woo 2003).
  
\item  The standard  model, assuming  adiabatic contraction  and \LCDM
  halo  concentrations, can  match the  $VL$ zero  point only  with an
  unrealistically  top-heavy IMF.   Even in  this case,  the predicted
  ratio  of $V_{2.2}/\Vvir$  is too  high for  a match  of  the galaxy
  luminosity function.

 \item Models  with a realistic  IMF and adiabatic  contraction require
  halo concentrations  that are a  factor two smaller  than predicted,
  which  is marginally  consistent with  the WMAP  constraints  on the
  cosmological  parameters.  Models with  realistic IMFs  and standard
  halo concentrations can simultaneously  match the $VL$ and $RL$ zero
  points only  if adiabatic  contraction does not  occur.  In  both of
  these cases the  predicted ratio of $V_{2.2}/\Vvir$ is  too high for
  matching the luminosity function.  
  
\item If disk formation causes dark-matter halos to expand rather than
  contract,  the  $VL$  and  $RL$  relations can  be  reproduced  with
  standard halo concentrations combined with low values of galaxy spin
  parameter  and mass  fraction.  The  low average  spin  parameter is
  consistent with  the picture  in which disks  survive in  halos that
  have  not  experienced recent  major  mergers  (D'Onghia \&  Burkert
  2004).   The low  baryon fraction  is consistent  with galaxy-galaxy
  lensing studies  (Mandelbaum \etal 2006).  This  model predicts that
  $V_{2.2} \simeq  1.2\Vvir$, consistent with what is  required to fit
  the luminosity function.
  
\item The scatter in the $VL$ relation has roughly equal contributions
  from  scatter in  $c$, in  $\YI$ and  from observational  errors. To
  match the  amount of scatter observed, we  require that $\sigma_{\ln
  c}\simeq0.23$.  This is smaller than the prediction for the full set
  of  dark-matter halos,  $\sigma_{\ln c}\simeq  0.32$  (Bullock \etal
  2001a), but  consistent with the sub-sample of  halos without recent
  major mergers (Wechsler \etal 2002).

\item The $RL$  scatter is dominated by scatter  in the spin parameter
  $\lamgal$.   The observed  $RL$ scatter  requires  that $\sigma_{\ln
    \lambda}  \simeq0.25$, about  half the  value predicted  for \LCDM
  halos.  This  is again  consistent with the  picture in  which disks
  survive in  halos with  a quiet merger  history.  This  picture also
  implies  that   the  {\it  average}   spin  parameter  is   low.   A
  simultaneous match  of the $VL$ and  $RL$ relations with  such a low
  $\overline{\lambda}_{\rm  gal}$  requires  that  the  baryonic  mass
  fraction is also  low.  This, in turn, strongly  favors a model with
  expansion rather than adiabatic contraction.

\item The observed residuals of  the $VL$ and $RL$ relations show only
  a  weak anti-correlation.   We have  shown that  this can  partly be
  attributed to the threshold density for star formation, which causes
  the spread in $\lamgal$ to  scatter galaxies along the $VL$ relation
  (see  also FA00).   Reproducing the  shallow slope  $\gamma$  of the
  residual correlation is possible only when there is scatter in $\YI$
  and when the scatter in  $c$ is relatively low, $\sigma_{\ln c} \lta
  0.23$.

\item Unlike the earlier suggestion by Courteau \& Rix (1999), we find
  that $\gamma$  does not  provide a strong  constraint on  the baryon
  fraction  in the  inner  halo.  The  relation  between $\gamma$  and
  $V_{\rm gal}/V_{2.2}$  depends on several model  ingredients such as
  the degree  of halo contraction or expansion,  the threshold density
  for star formation,  and the sensitivity of $\gamma$  to the scatter
  in the different variables.  In addition, in all our models there is
  a  large   scatter  in  $V_{\rm  gal}/V_{2.2}$   which  is  strongly
  correlated with surface brightness.  This correlation results in the
  highest  surface brightness  galaxies being  baryon within  2.2 disk
  scale lengths,  while at the  same time allowing the  lowest surface
  brightness  galaxies to  be dark  matter dominated  as  is generally
  accepted.

\item  Although  observations  have  shown  that  the  $VL$  and  $RL$
  residuals  are uncorrelated,  our models  robustly predict  that the
  residuals of the {\it baryonic} $VM$ and $RM$ relations are strongly
  anti-correlated.

\end{itemize}
Based  on these  results we  conclude that  the standard  model, which
includes standard  adiabatic contraction, standard  halo concentration
parameters and a standard IMF,  does not allow a simultaneous match to
the  $VL$ and  $RL$  relations.   This is  a  modification of  earlier
conclusions, e.g., by MMW and  Pizagno \etal (2005), which owes to our
more  realistic modeling  (see Appendix~\ref{sec:Piz05}).   Although a
proper   fit  can  be   obtained  if   the  halo   concentrations  are
significantly lower, or if adiabatic contraction does not occur, these
models predict  high values of  $V_{2.2}/\Vvir$ which prevent  a match
with the galaxy luminosity  function.  They also predict high baryonic
mass fractions in conflict with results from galaxy-galaxy lensing.

To circumvent  these problems, we advocate  a model in  which the dark
matter halo responds  by expansion rather than by  contraction, to the
formation  of   the  disk.   This   model  (i)  predicts   values  for
$V_{2.2}/\Vvir$  that allow  a  simultaneous  fit to  the  LF (ii)  is
consistent with disks forming  predominantly in dark matter halos that
have  not  experienced  a  recent  major merger,  (iii)  predicts  low
baryonic  mass   fractions,  in  agreement   with  a  wide   range  of
observations,  and  (iv) predicts  dark  matter  halos  that are  less
centrally concentrated, in better agreement with rotation curve shapes
and bar pattern speeds.

The idea of  an expanding dark matter halo is  counter to the standard
model for disk formation, in which disks form out of relatively smooth
cooling flows  that conserve their specific  angular momentum. Rather,
the expansion scenario requires that disks form out of merging clumps,
which  transfer energy  and angular  momentum to  the dark  matter via
dynamical  friction  and three  body  interactions.   Such a  scenario
naturally arises  if disks  form out of  clumpy, cold  accretion flows
(Birnboim \&  Dekel 2003; Keres  \etal 2005; Dekel \&  Birnboim 2006).
While  the formation  of realistic  disk galaxies  out of  such clumpy
streams is  yet to  be investigated in  detail, preliminary  hints are
provided  by hydrodynamical simulations  which have  demonstrated that
disks can originate from  merging gaseous progenitors (e.g., Robertson
\etal 2006).

\smallskip We thank E. Bell,  S.M.  Faber, A.  Maller \& B.  Robertson
for stimulating discussions.  A.A.D.  has been partly supported by the
Swiss  National  Science  Foundation  (SNF).  A.D.  has  been  partly
supported by ISF 213/02, NASA ATP NAG5-8218, a Miller Professorship at
UC Berkeley, and  a Blaise Pascal International Chair  in Paris.  S.C.
acknowledges the support of NSERC through a Discovery grant.



\clearpage

\begin{appendix}

\section{A:  Comparison with Pizagno \etal 2005}
\label{sec:Piz05}
Pizagno  \etal  (2005, hereafter  P05)  study  the correlations  among
stellar mass, $M_{*}$, $I$-band disk scale length, $\RI$, and rotation
velocity at 2.2 disk scale lengths, $V_{2.2}$, for a sample of 81 disk
dominated galaxies (defined to have disk-to-total luminosity fractions
greater than 0.9).  In what  follows we refer to the $V_{2.2}-M_*$ and
$\RI-M_*$ relations as the $VM$ and $RM$ relations, respectively.  P05
estimate stellar  masses using  $g-r$ colors corrected  for extinction
and the relations  in Bell \etal (2003a).  As  shown in \S\ref{sec:ml}
the $\YI  -\LI$ relation  of P05 is  in excellent agreement  with ours
(assuming the same IMF).

P05 claim that a MMW  model with $\md=0.05, \lam=0.06, c_{200}=10$ and
using the adiabatic contraction formula of Gnedin \etal (2004) matches
the $VM$ and $RM$ data  reasonably well.  This is in disagreement with
our statement that the MMW model  is unable to reproduce the slopes of
the $VL$  and $RL$ relations  (\S\ref{sec:slopes}) and with  our claim
that models with  adiabatic contraction, standard halo concentrations,
and  standard IMFs require  $\mgal\simeq\lamgal\simeq0.1$ in  order to
match  the $VL$  zero point  (Fig.~\ref{fig:chi5}).  Below  we address
these and other differences between our results and those of P05.

\begin{itemize}  
\item  We  first note  that  unlike our  $VL$  relation  and those  of
  Giovanelli \etal  1997, the  $VM$ and $VL$  relations of P05  do not
  follow a  single power-law. There  is a significant  deviation below
  $V_{2.2}=120 \kms$ in  both their $VM$ and $VL$  relations.  In what
  follows we focus on their galaxies with $V_{2.2}>120\kms$.

\item P05 assume  $c_{200}=10$ and that the disk  is 100\% stars, thus
  their $VM$  relation has  a slope of  $1/3$, in agreement  with their
  data.   However,  as   shown  in  \S\ref{sec:slopes},  the  expected
  variation of $c$ and the gas-to-stellar mass ratio with $\Mvir$ both
  result in  significantly shallower $VM$ slopes. Thus  either $c$ and
  the  gas-to-stellar  mass  ratio  do  not  depend  significantly  on
  $\Mvir$, or their data significantly over-estimates the $VM$ slope.

\item  As can  clearly be  seen in  Fig.~4 of  P05, their  model with
  $\md=0.05$ and  $\lam=0.06$ does  not match the  $VM$ zero  point of
  the data.  Their model  with $\md=0.10$ and $\lam=0.08$ provides a
  better  match,  but predicts  a  much  stronger  dependence of  $VM$
  scatter on $\lambda$, and hence  disk size.  Their models assume
  pure stellar  disks.   Including reasonable  gas-to-stellar  mass
  ratios requires even  larger values of $\md$ and  $\lambda$ to match
  the $VM$ zero  point which are now consistent  with our results (see
  Fig.~\ref{fig:chi5}).

\item P05  do not construct a  self consistent model for  the $VM$ and
  $RM$ relations and  their scatter.  We have verified  that pure disk
  models (e.g. MMW) which simultaneously match  the $VM$  and $RM$
  relations  {\it always}  predict a  significant  correlation between
  $VM$  and $RM$ residuals.   P05 speculated  that a  weak correlation
  between size and $VM$ residual  could be washed out by other sources
  that contribute  to the $VM$ scatter.  However  the only significant
  sources to  the $VM$  scatter are $c$  and observational  errors. We
  have found that while observational errors do reduce the strength of
  the residual correlation the effect is small. By contrast scatter in
  $c$ results in a significantly stronger negative correlation between
  the residuals.  Thus rather than washing out the correlation between
  size   and  $VM$  residual,   the  expected   scatter  in   $c$  and
  observational errors will result in a stronger correlation.
  
\item    The     $RL$    relation     of    P05,    as     shown    in
  Fig.~\ref{fig:VLR_ALL_lit},  has a  significantly higher  zero point
  normalization than ours.  The most likely sources of this difference
  are their  bulge-to-disk ratio selection criteria, or  their lack of
  inclination corrections to the  disk sizes.  Their larger disk sizes
  require a $\simeq 50  \%$ larger spin parameter, $\lamgal$, assuming
  all the other  model parameters are kept fixed.   Larger disks, at a
  given disk  mass, contribute less  to $V_{2.2}$, and result  in less
  halo contraction.  This  makes it easier to fit  the $VL$ zero point
  and  also  weakens  the   correlation  between  the  $VL$  and  $RL$
  residuals.  However,  as can be  seen in Fig.~\ref{fig:chi5}  even a
  50\%  increase  in  the  $RL$  zero point  (corresponding  to  about
  1$\sigma$  of  the observational  scatter)  would not  significantly
  change any of our conclusions  regarding a simultaneous match the of
  the $VL$ and $RL$ zero points and a low ratio of $V_{2.2}/\Vvir$.

\end{itemize}

In summary, although there are differences in the data samples used by
P05  and  ourselves,   both  our  and  the  P05   data  require  large
$\mgal\simeq\lamgal\simeq0.1$ in  order to match the $VL$  zero point. 
Not only  is such a high  value of $\lamgal$ unrealistic,  such a high
$\mgal$ is  also inconsistent with galaxy-galaxy  lensing constraints. 
Furthermore the  MMW model adopted by  P05 is unable  to reproduce the
slope or the surface brightness independence of the $VL$ relation, due
to the overly simplistic assumption that galaxy disks are 100\% stars.
 
\section{B: Correlations between model parameters}
\label{sec:modcor}

We  have  assumed  that  the  scatter  in  the  model  parameters  are
uncorrelated. For completeness, we  here present a brief discussion on
how correlations between model parameters might affect our results.

\begin{itemize}
  
\item $\lambda$-$c$  anti-correlation: Bullock \etal  (2001b) found no
  correlation  between $\lambda$  and $c$  above the  weak correlation
  expected from  the definition  of $\lambda$.  However,  Bailin \etal
  (2005a)  claim  a correlation,  but  this  is  likely due  to  their
  inclusion of unrelaxed halos (Macci{\`o} \etal 2006).  Suppose halos
  with    larger    $\lambda$    have    lower   $c$;    looking    at
  Fig.~\ref{fig:VLR-mcl8}  we see  that  this would  result in  larger
  scatter  in  both the  $RL$  and $VL$  relations,  and  by the  same
  reasoning a stronger residual correlation.
  
\item  $\YI-c$ correlation:  At a  fixed  halo mass,  lower $c$  halos
  assemble later; in addition lower  $c$ halos result in lower surface
  density disks. Thus, everything else  being equal, we expect that at
  a fixed halo mass, lower  $c$ halos to contain galaxies with younger
  (bluer)  stellar  populations than  halos  with  larger $c$.   Since
  younger  populations  correspond   to  lower  stellar  mass-to-light
  ratios, we thus expect a positive correlation between $c$ and $\YI$.
  From  Fig.~\ref{fig:VLR-mcl8}   we  see  that   such  a  correlation
  increases the overall $VL$ scatter.   It will also decrease the $RL$
  scatter, but since $\YI$ and $c$ contribute relatively little to the
  $RL$ scatter, this reduction is unlikely to be significant.
  
\item  $\lamgal-\mgal$  correlation:  since  mass  is  more  centrally
  concentrated  than specific angular  momentum one  generally expects
  that $\jgal < \mgal$ if disk galaxies form inside out.  This implies
  that a  smaller $\mgal$ implies a smaller $\lamgal$.   If these
  two parameters are indeed positively correlated, the overall scatter
  in the $RL$ relation is predicted to be smaller, as can be seen from
  Fig.~\ref{fig:VLR-mcl8}.  However, recall (\S 4.1 \& 4.3) that it is
  the correlation between the gas-to-stellar mass ratio and $\lamgal$,
  through  a  critical  star  formation  threshold  density,  that  is
  essential  in reproducing  the slope  of the  $VL$ relation  and the
  surface brightness  independence of  the $VL$ relation.   Any models
  that  introduce a  correlation  between $\lamgal$  and $\mgal$  will
  likely erase these successes.
    
\item  $\Upsilon-\lamgal$ anti-correlation: At  a fixed  baryonic mass
  lower  $\lamgal$ results  in  higher surface  density disks.   Since
  empirically  the star  formation efficiency  is proportional  to the
  surface density of the stars (Kennicutt 1998), we thus expect higher
  surface  density  systems  to  have older  stellar  populations,  in
  agreement  with   the  data   (Kauffmann  \etal  2003b).    Such  an
  anti-correlation would  tend to remove  any color dependence  of the
  $RL$  relation and  hence reduce  its  scatter (Bell  \etal 2003b).  
  However, it would also increase the surface brightness dependence of
  the $VL$ relation.

\end{itemize}

In  summary, there  are several  plausible correlations  between model
parameters. Although three of these  have the potential to explain why
the  $RL$ scatter is  smaller than  predicted, we  do not  expect that
these  correlations  can  reconcile  the  full amount  of  scatter  in
$\lambda$ expected  from simulations with the observed  scatter in the
$RL$ relation.  Furthermore, each  of these correlations will increase
the strength  of the correlation  between $VL$ and $RL$  residuals (or
equivalently the strength of  the surface brightness dependence of the
$VL$ relation).  Thus to  reproduce the weakly correlated residuals we
would require even less scatter  in $c$ than the $\sigma_{\ln c}=0.23$
we currently adopt. This  reinforces the conclusion that disk galaxies
form in a subset of halos.

\section{C: Courteau \& Rix (1999) revisited}
\label{sec:CR99}

\begin{figure*}
\begin{center}
\includegraphics[width=6.0in]{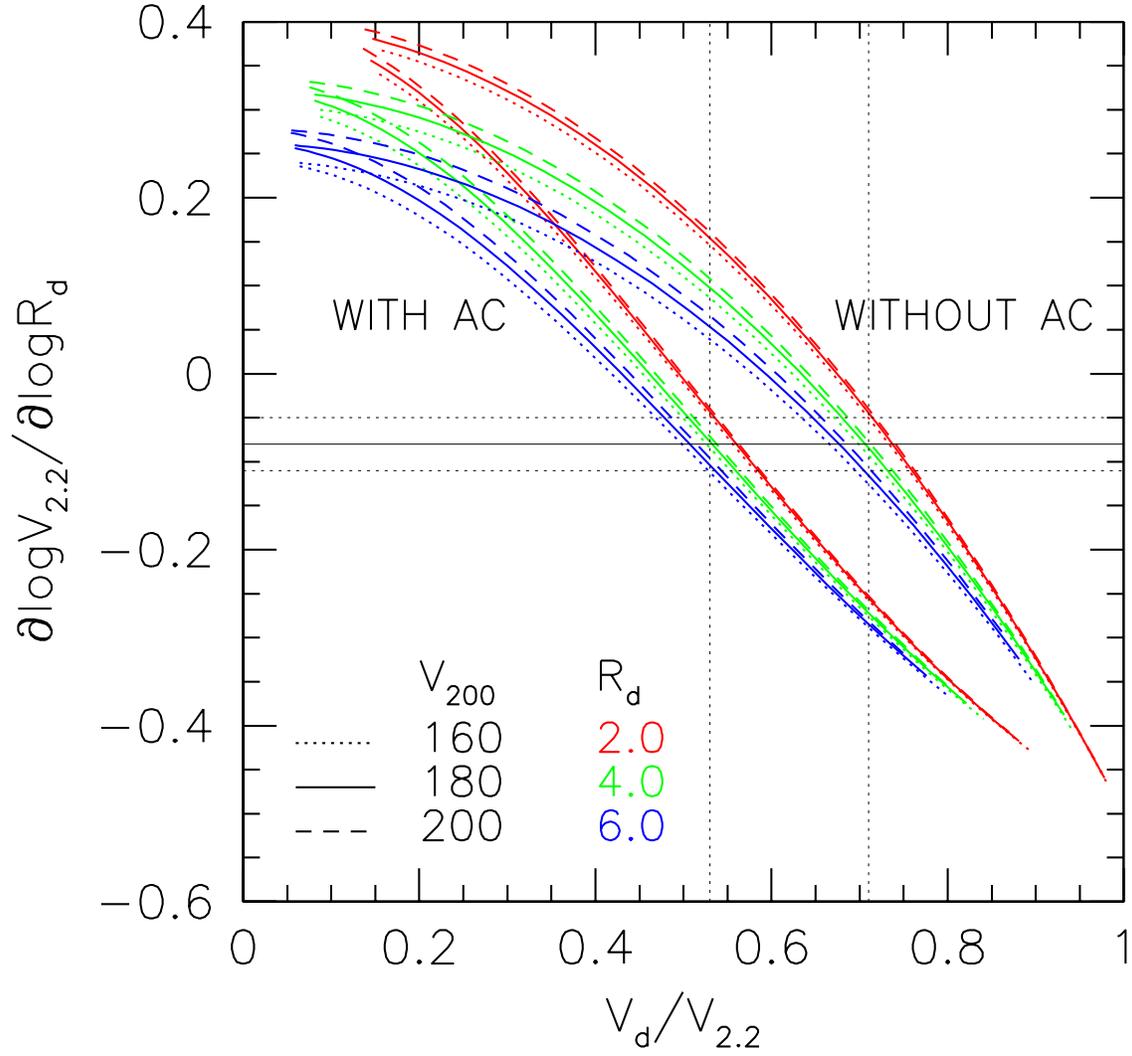}
\caption{
  CR99  analysis for  NFW  halos with  $c_{200}=10$  with and  without
  adiabatic contraction  (AC).  Note that  $c_{vir}\simeq 1.3c_{200}$. 
  The $y$-axis  shows the change in  $\log V_{2.2}$ due  to changes in
  $\log\Rd$ at  fixed $\Md$.  The  line and color types  correspond to
  different values of $V_{200}$  and $\Rd$, respectively, as indicated
  in  the figure.   The horizontal  solid  and dotted  lines show  the
  observed  slope of  the $\Delta\log  V-\Delta\log R$  relation.  The
  vertical dotted lines show  the $V_{\rm d}/V_{2.2}$ corresponding to
  the  $V_{200}=180,   \Rd=4$  models.   Models  without   AC  have  a
  substantially  larger $V_{\rm d}/V_{2.2}$  (about 30\%)  than models
  with AC. }
\label{fig:CR99_c10}
\end{center}
\end{figure*}

CR99 argued  that the  weak correlation between  the residuals  of the
$VL$  and $RL$  relations, $\gamma\simeq-0.1$,  implies disks  of high
surface brightness galaxies should  have, {\it on average} $\vdt\simeq
0.6$.  CR99 arrive at this result by computing $\partial\log V_{2.2} /
\partial \log \Rd$  for models consisting of exponential  disks in NFW
halos  with adiabatic contraction.   They found  that in  these models
\dvr correlates with $\vdt$ (the ratio of the circular velocity of the
disk to the total circular  velocity, at 2.2 disk scale lengths), such
that galaxies with  higher $\vdt$ have more negative  $\dvr$, with the
limiting case  of $\dvr=-0.5$  for a pure  exponential disk.   Thus by
assuming  that $\dvr$  is  equivalent  to the  observed  slope of  the
residual correlation  $\gamma$, they concluded that  $\vdt\simeq 0.6$. 
This would suggest more DM than baryons within 2.2 disk scale lengths,
but not that the baryon contribution is insignificant.  If the baryons
were  insignificant then  the CR99  model would  predict $\dvr  >  0$. 
However, as  outlined below, several  of the assumptions made  in CR99
are not true in general.

\begin{itemize}
\item CR99 assumed that  halos contract adiabatically to the formation
  of the  disk.  In Fig.~\ref{fig:CR99_c10}, which  should be compared
  with Fig.~9 of CR99, we repeat  the CR99 analysis for halos with and
  without adiabatic contraction.  Note that CR99 used $c_{200}=10$ and
  that $c_{\rm vir}\simeq  1.3 \, c_{200}$, for ease  of comparison we
  also adopt their definition here.  We see that turning off adiabatic
  contraction increases $\vdt$ by about $30\%$ at a fixed $\dvr$.

\item  CR99 assumed  that bulge  formation does  not affect  the scale
  length of the disk. We have  shown that in models where the specific
  angular momentum of the disk  increases due to bulge formation (e.g. 
  $\fx=0.25$) the residual correlation is weakened.
  
\item CR99  assumed that the  contribution of the  gas to the  disk is
  negligible.  While  this is true for the  highest surface brightness
  galaxies, moderate surface brightness galaxies contain more gas.  We
  have  shown (see \S\ref{sec:rescorr})  that the  correlation between
  the gas-to-stellar mass ratio and disk surface density significantly
  reduces the strength of the residual correlations.
  
\item CR99 underplayed the effect of scatter in halo concentration $c$
  and  stellar  mass-to-light  ratio $\Upsilon$.  Fig.~\ref{fig:dVLR3}
  shows that  scatter in these parameters can  result in significantly
  different residual  correlation slopes. Scatter in $c$  results in a
  more negative correlation  while scatter in $\YI$ results  in a more
  positive correlation.  For realistic amounts of scatter  in both $c$
  and $\YI$ the residual correlation is weakened.
  
\item CR99 assumed that $\gamma  \equiv \dvr$.  In general this is not
  true, because $\gamma$ is a  global quantity while $\dvr$ is a local
  quantity. More  specifically $\gamma$ is  the slope of the  $VL$ and
  $RL$ residuals for  a sample of galaxies, while  $\dvr$ is the slope
  of the change in $V$ due to change in $R$, at fixed disk mass, for a
  single galaxy.

\end{itemize}

Thus we conclude that the  $VL$-$RL$ residuals cannot be used to place
model  independent  constraints  on  the  baryonic  fraction  of  disk
galaxies.   However, this  does  not diminish  their  importance as  a
constraint for galaxy formation models.


\end{appendix}

\end{document}